\documentclass[aps,pra,twocolumn,superscriptaddress,showpacs,nofootinbibfloatfix,amsmath,amsfonts,amssymb]{revtex4-2}%
\usepackage{amsmath,amsfonts,amssymb,color}
\usepackage{amsthm}
\usepackage{leftidx}
\usepackage{graphicx}
\usepackage{xcolor}
\usepackage{dcolumn}
\usepackage{bm}
\usepackage{epstopdf}
\usepackage{epsfig}
\usepackage{environ}
\usepackage{pdfcomment}

\usepackage{multirow}
\usepackage{setspace}
\usepackage{color}

\usepackage{float}
\usepackage[T1]{fontenc}
\usepackage[latin9]{inputenc}
\usepackage{setspace}
\usepackage{esint}

\begin{document}


\title{Quantum geometry and geometric entanglement entropy of one-dimensional
	Floquet topological matter}

\author{Longwen Zhou}
\email{zhoulw13@u.nus.edu}
\affiliation{%
	College of Physics and Optoelectronic Engineering, Ocean University of China, Qingdao, China 266100
}
\affiliation{%
	Key Laboratory of Optics and Optoelectronics, Qingdao, China 266100
}
\affiliation{%
	Engineering Research Center of Advanced Marine Physical Instruments and Equipment of MOE, Qingdao, China 266100
}

\date{\today}

\begin{abstract}
The geometry of quantum states could offer indispensable insights
for characterizing the topological properties, phase transitions and
entanglement nature of many-body systems. In this work, we reveal
the quantum geometry and the associated entanglement entropy (EE)
of Floquet topological states in one-dimensional periodically driven
systems. The quantum metric tensors of Floquet states are found to
show non-analytic signatures at topological phase transition points.
Away from the transition points, the bipartite geometric EE of Floquet
states exhibits an area-law scaling vs the system size, which holds
for a Floquet band at any filling fractions. For a uniformly filled
Floquet band, the EE further becomes purely quantum geometric. At
phase transition points, the geometric EE scales logarithmically with
the system size and displays cusps in the nearby parameter ranges.
These discoveries are demonstrated by investigating typical Floquet
models including periodically driven spin chains, Floquet topological
insulators and superconductors. Our findings uncover the rich quantum
geometries of Floquet states, unveiling the geometric origin of EE
for gapped Floquet topological phases, and introducing information-theoretic
means of depicting topological transitions in Floquet systems.
\end{abstract}

\maketitle

\section{Introduction\label{sec:Int}}

Floquet topological phases have attracted sustained research interest
over the last decades \cite{FTPRev01,FTPRev02,FTPRev03,FTPRev04,FTPRev05,FTPRev06,FTPRev07,FTPRev08,FTPRev09}.
It was found that periodic driving fields could endue a system with
rich and unique features that are absent or challenging to achieve
in static systems, such as Floquet phases with large topological invariants
and many topological edge states \cite{HoPRL2012}, dispersionless
Floquet edge modes at quasienergy $\pi$ \cite{JiangPRL2011} or anomalous
chiral edge modes encircling the quasienergy Brillouin zone (BZ) \cite{RudnerPRX2013},
and exotic phenomena like Floquet-band holonomy \cite{LWZPRB2016}
and integer quantum Hall effect from chaos \cite{TianPRL2014}. Experimental
realizations of these intriguing physics in both solid-state materials
and quantum simulators \cite{FTPExp01,FTPExp02,FTPExp03,FTPExp04,FTPExp05,FTPExp06,FTPExp08,FTPExp09,FTPExp10,FTPExp11,FTPExp12,FTPExp13,FTPExp14,FTPExp15,FTPExp16,FTPExp17}
brought about their potential applications in topological photonic
devices \cite{BlochNat2022,GaliffiAP2022,YinELight2022}, ultrafast
electronics \cite{FTPRev04,FTPRev08} and novel quantum computing
strategies \cite{BomantaraPRL2018,BomantaraPRB2020}.

In contrast with topological aspects, less attention was paid
to the geometric properties of Floquet states \cite{WeinbergPR2017}
and their resulting entanglement characteristics \cite{LWZPRR2022}.
The geometries of quantum states, including the amplitude and phase
distances described respectively by the quantum metric and Berry curvature
tensors \cite{TormaPRL2023}, have played pivotal roles in the study
of Bloch-electron dynamics \cite{XiaoRMP2010}, topological states
of matter \cite{VanderbiltBook2018} and quantum phase transitions
\cite{ZhuIJoMPB2008,GuIJoMPB2010,CarolloPR2020}. For example, the
integration of Berry curvature over a two-dimensional BZ
yields the first Chern number of a Bloch band, which serves as the
topological origin of various transport phenomena including the integer
quantum Hall effect \cite{TKNN1982}, quantum anomalous Hall effect
\cite{Haldane1988} and quantized adiabatic charge pumping \cite{ThoulessPump1983}.
The quantum metric tensor could instead appear in the high-order response
coefficient of electrons to external fields \cite{NLQHENat2023} and
the superfluid weight of flat bands \cite{PeottaNC2015}, yielding
important insights for the understanding of nonlinear Hall effects
\cite{NLQHESci2023} and flat-band superconductivity in correlated
materials \cite{TianNat2023}. In periodically driven systems, the
quantum geometry of Floquet bands may also offer essential information
about the topological and entanglement features of the underlying
nonequilibrium states. First, as Floquet bands could weave
around the first quasienergy BZ $E\in[-\pi,\pi)$, they may develop
level crossings at both the quasienergies zero and $\pi$, yielding
two possible flavors of topological phase transitions \cite{JiangPRL2011}.
Whether and how these transitions would leave unique signals in the
quantum geometric tensor of Floquet states then constitute interesting
issues to address. Second, driving fields could generate long-range
couplings in a system and allow Floquet bands to carry large topological
invariants \cite{HoPRL2012}. Unveiling geometric aspects of these
quasienergy bands with large topological numbers may help us to deepen
our understanding of the quantum transport in driven systems. Third,
Floquet systems could possess anomalous topological phases with unique
edge states, such as degenerate edge modes at the quasienergy $\pi$
\cite{RudnerPRX2013}, which are not reachable in static settings.
Geometric signatures of these anomalous Floquet topological phases
deserve to be further clarified. Resolving these issues
thus forms an indispensable part for our understanding of Floquet
topological matter and their entanglement properties.

In this paper, we uncover the quantum geometries of Floquet-Bloch
bands and their associated entanglement nature in one-dimensional
(1D) Floquet topological phases. In Sec.~\ref{sec:QGEE}, we outline
the generic definitions of Abelian quantum geometric tensor and geometric
entanglement entropy (GEE) of Floquet states, with further theoretical
details presented in Appendices \ref{sec:Ovlp}--\ref{sec:GEE}.
Based on these definitions, we obtain the quantum metric tensor (QMT)
and GEE of typical 1D Floquet systems including periodically driven
spin chains, Floquet topological insulators and superconductors in
Secs.~\ref{sec:HDSC}--\ref{sec:PQKC}. Throughout these model studies,
we reveal that the integrated QMT of a filled Floquet band would show
non-analytic signatures when the system undergoes a transition between
different Floquet topological phases. Moreover, away from the
transition point, the GEE always follow an area-law scaling vs the
system size irrespective of the filling fraction of the considered
Floquet band, and the bipartite EE becomes purely geometric when the
Floquet band is uniformly filled. At the transition point between
different Floquet topological phases, the EE is also of geometric
origin and further follows a log-law scaling vs the system size, as
expected in 1D critical metallic phases. In Sec.~\ref{sec:Sum}, we
summarize our results and discuss potential directions of future research.

\section{Quantum geometry and geometric entanglement entropy\label{sec:QGEE}}

In this section, we outline the definitions of key quantum geometric
objects and their related entanglement measures that will be investigated
in this study. Further derivation details of these quantities are
given in the Appendices \ref{sec:Ovlp}--\ref{sec:GEE}.

Consider a set of normalized quantum states $\{|\psi({\bf k})\rangle\}$,
which are defined in a continuous $D$-dimensional parameter space
${\bf k}=(k_{1},k_{2},...,k_{D})$. The infinitesimal distance between
any two nearby states in such a ${\bf k}$-space can be expressed
as $ds^{2}\equiv1-||\langle\psi({\bf k})|\psi({\bf k}+d{\bf k})\rangle||^{2}$,
which is equal to one (zero) if the states $|\psi({\bf k})\rangle$
and $|\psi({\bf k}+d{\bf k})\rangle$ are orthogonal (identical up
to a phase factor). Retaining terms up to the second-order in $d{\bf k}$,
we can equivalently write $ds^{2}$ as
\begin{equation}
	ds^{2}={\rm Re}\left[{\cal Q}_{\alpha\beta}({\bf k})\right]dk_{\alpha}dk_{\beta}=g_{\alpha\beta}({\bf k})dk_{\alpha}dk_{\beta},\label{eq:ds2}
\end{equation}
where the indices $\alpha,\beta=1,2,...,D$ are summed over. The quantity
${\cal Q}_{\alpha\beta}({\bf k})$, given by
\begin{equation}
	{\cal Q}_{\alpha\beta}({\bf k})=\langle\partial_{k_{\alpha}}\psi({\bf k})|\left[1-|\psi({\bf k})\rangle\langle\psi({\bf k})|\right]|\partial_{k_{\beta}}\psi({\bf k})\rangle,\label{eq:QGT}
\end{equation}
is usually referred to as the component of quantum geometric tensor
(QGT). The real part of QGT gives the QMT
\cite{QGT1980}, whose components are given by the $g_{\alpha\beta}({\bf k})$
in Eq.~(\ref{eq:ds2}). The integration of QMT over ${\bf k}$-space
may provide further diagnoses for level crossings and quantum phase
transitions in the system \cite{ZanardiPRL2007}. The imaginary part
of QGT yields the Berry curvature ${\cal F}({\bf k})$ \cite{Berry1984},
whose components are
\begin{equation}
	{\cal F}_{\alpha\beta}({\bf k})=-2{\rm Im}\left[{\cal Q}_{\alpha\beta}({\bf k})\right].\label{eq:BC}
\end{equation}
The Berry curvature of Bloch bands determines the anomalous dynamics
and quantized Hall response of electrons in two-dimensional systems
\cite{XiaoRMP2010}. The integration of ${\cal F}_{\alpha\beta}({\bf k})$
over a closed and orientable two-dimensional ${\bf k}$-space manifold
further yields the first Chern number, which is a key ingredient in
characterizing topological phases of matter \cite{VanderbiltBook2018}.
The information provided by $g_{\alpha\beta}({\bf k})$ and ${\cal F}_{\alpha\beta}({\bf k})$
thus offers a complete description for the geometry of quantum states
$\{|\psi({\bf k})\rangle\}$ in ${\bf k}$-space. Note in passing
that for a 1D ${\bf k}$-space, the Berry curvature
vanishes by definition and the QGT reduces to a one-component QMT,
i.e.,
\begin{equation}
	g_{kk}=\langle\partial_{k}\psi(k)|\left[1-|\psi(k)\rangle\langle\psi(k)|\right]|\partial_{k}\psi(k)\rangle.\label{eq:gkk0}
\end{equation}
In Appendix \ref{sec:QMT}, the expressions of $g_{kk}$ for generic
1D two-band models and for some representative examples are worked
out explicitly, with $k$ being identified as the 1D quasimomentum defined in the
first BZ $[-\pi,\pi)$. These expressions will be used
to characterize the quantum geometry of Floquet-Bloch bands in later
sections.

Quantum entanglement comprises the non-classical correlations among
different constituents of a composite quantum system. Related information-theoretic
measures, such as the entanglement spectrum and EE, have been regularly adopted in depicting quantum phase transitions
and topological phases in many-body systems (for reviews see \cite{ESEERev01,ESEERev02,ESEERev03,ESEERev04,ESEERev05,ESEERev06,ESEERev07,ESEERev08,ESEERev09,ESEERev10,ESEERev11}).
In a recent study, it was found that the geometry of quantum states
could contribute a universal area-law component to the bipartite EE
of noninteracting fermions in static multi-band models \cite{PaulPRB2024}.
Such a geometric entanglement entropy may be defined as
\begin{equation}
	S_{{\rm QG}}\equiv S_{A}-S_{A_{0}},\label{eq:SQG}
\end{equation}
where $S_{A}$ is the (R\'enyi or von Neumann) EE between the subsystem
$A$ and its complementary $\overline{A}$ in a bipartite system $A\cup\overline{A}$
with fermions populating a Bloch band. $S_{A_{0}}$ encompasses
the bipartite EE of fermions sharing the same Fermi surface with those
in the system $A\cup\overline{A}$ but with trivial Bloch band geometries.
The difference between $S_{A}$ and $S_{A_{0}}$ then yields an entropic
component originated from the inherent quantum geometry of occupied
Bloch states \cite{PaulPRB2024}.

For electrons in a 1D periodic lattice with $L$ unit cells, the $S_{A_{0}}$
can be obtained rather generally from the spectrum of overlap matrix
$O^{A_{0}}$ among plane-wave basis $\{\langle n|k\rangle=L^{-1/2}e^{ikn}|n=1,...,L\}$,
whose matrix elements are given by
\begin{equation}
	O_{k,k'}^{A_{0}}=\frac{1}{L}\sum_{n\in A}e^{-i(k-k')n}.\label{eq:OA0kkp0}
\end{equation}
Here $L$ is the total number of unit cells in the composite system
$A\cup\overline{A}$, and the cell index $n$ has been restricted
to the subsystem $A$. $k$ and $k'$ are wave vectors running over
all the occupied single-particle eigenbasis, so that $O^{A_{0}}$
is an $N\times N$ matrix if there are $N$ particles in the system
$A\cup\overline{A}$. Denoting the eigenvalues of $O^{A_{0}}$ by
$\{\eta_{\ell0}|\ell=1,...,N\}$, we can obtain the von Neumann EE
$S_{A_{0}}$ in Eq.~(\ref{eq:SQG}) as \cite{Fredholm01,Fredholm02,Fredholm03}
\begin{equation}
	S_{A_{0}}=-\sum_{\ell=1}^{N}\left[\eta_{\ell0}\ln\eta_{\ell0}+(1-\eta_{\ell0})\ln(1-\eta_{\ell0})\right],\label{eq:SA0}
\end{equation}
with further derivation details presented in the Appendices \ref{sec:EE}--\ref{sec:GEE}.
Meanwhile, if $\{|\psi_{k}\rangle\}$ constitutes the occupied eigenstates
of a Bloch band in the system $A\cup\overline{A}$, we can construct
an overlap matrix $O^{A}$ among the states in $\{|\psi_{k}\rangle\}$.
Its matrix elements (after being restricted to the subsystem $A$)
are given by
\begin{equation}
	O_{k,k'}^{A}=\frac{1}{L}\sum_{n\in A}e^{-i(k-k')n}\langle\psi_{k}|\psi_{k'}\rangle.\label{eq:OAkkpmain}
\end{equation}
With $N$ particles in the system, $O^{A}$ is also $N\times N$ with
$N$ eigenvalues $\{\eta_{\ell}|\ell=1,...,N\}$, from which the von
Neumann bipartite EE in Eq.~(\ref{eq:SQG}) can be obtained as \cite{Fredholm01,Fredholm02,Fredholm03}
\begin{equation}
	S_{A}=-\sum_{\ell=1}^{N}\left[\eta_{\ell}\ln\eta_{\ell}+(1-\eta_{\ell})\ln(1-\eta_{\ell})\right].\label{eq:SA}
\end{equation}
Further derivation details of $S_{A}$ can also be found in the Appendices
\ref{sec:EE}--\ref{sec:GEE}. Note in passing that formally speaking,
both the $S_{A_{0}}$ and $S_{A}$ in Eqs.~(\ref{eq:SA0}) and (\ref{eq:SA})
do not explicitly depend on the size of subsystem $A$. Within a Bloch
band, all the nontrivial quantum geometries of the occupied states
$\{|\psi_{k}\rangle\}$ are encoded in their overlaps $\langle\psi_{k}|\psi_{k'}\rangle$
with $k\neq k'$. We thus expect that the difference between $S_{A}$
and $S_{A_{0}}$ in Eq.~(\ref{eq:SQG}) could properly describe a
quantum geometric component of EE, as the trivial contribution $S_{A_{0}}$
(from a free fermion gas in a periodic lattice without onsite potentials)
has been removed (see Appendices \ref{sec:EE}--\ref{sec:GEE} from
further discussions). It was also revealed that for fermions filling
a gapped Bloch band, both $S_{A}$ and $S_{A_{0}}$ scales as $L_{A}^{D-1}\ln L_{A}$
vs the subsystem size $L_{A}$ up to its leading order, with the same
scaling coefficient in $D$ spatial dimensions \cite{GioevPRL2006}.
Their difference should thus follow an area-law scaling vs $L_{A}$
when the system is away from its critical point. One may then identify
quantum and topological phase transitions from the change of finite-size
scaling behaviors in $S_{{\rm QG}}$.

A Floquet quantum system can be described by a Hamiltonian $\hat{H}(t)=\hat{H}(t+T)$,
which is periodic in time $t$ with the driving period $T$. If we
are interested in the stroboscopic dynamics of the system, we can
focus on its Floquet operator $\hat{U}=\hat{\mathsf{T}}e^{-i\int_{t}^{t+T}\hat{H}(t')dt'}$,
which controls the evolution of the system over a complete driving
period ($\hat{\mathsf{T}}$ performs the time ordering). The eigenvectors
and eigenphases of $\hat{U}$ are respectively called the Floquet
eigenstates and quasienergies, which could be obtained by solving
the eigenvalue equation $\hat{U}|\psi\rangle=e^{-iE}|\psi\rangle$.
The Floquet eigenstates form a complete and orthonormal basis of the
system. If $\hat{H}(t)$ also possesses some discrete spatial translational
symmetries, $\hat{U}$ will hold the same symmetries. Its quasienergies
could then be grouped into bands confined in the first quasienergy
BZ $E\in[-\pi,\pi)$, which are called the Floquet-Bloch
bands. In this case, a Floquet eigenstate $|\psi_{j}({\bf k})\rangle$
in the quasienergy band $E_{j}({\bf k})$ satisfies the equation $\hat{U}|\psi_{j}({\bf k})\rangle=e^{-iE_{j}({\bf k})}|\psi_{j}({\bf k})\rangle$,
with $j$ the band index and ${\bf k}$ the quasimomentum. In Appendices
\ref{sec:EE}--\ref{sec:GEE}, we demonstrate that for fermions
filling the Floquet-Bloch band of a 1D periodically driven system,
the formalism of QMT and GEE as outlined in this section are also applicable
after the replacement of each filled Bloch state with a Floquet-Bloch
eigenstate at a given quasienergy in the related equations. This allows
us to unveil the quantum geometry and the associated EE of some representative
1D periodically driven systems, including Floquet spin chains, topological
insulators and superconductors in the following sections.

\section{Harmonically driven spin chain: QMT and GEE\label{sec:HDSC}}

We start with a ``minimal'' Floquet model, whose geometric, topological
and entanglement properties could be controlled by periodic driving
fields. The model describes a 1D spin chain subject to harmonic drivings
\cite{YangPRB2019}, whose Hamiltonian takes the form
\begin{widetext}
\begin{equation}
	\hat{H}(t)=\sum_{n}\left\{ \frac{\delta_{1}[1-\sin(\omega t)]}{4}\hat{\sigma}_{n}^{x}\hat{\sigma}_{n+1}^{x}+\frac{\delta_{1}[1+\sin(\omega t)]}{4}\hat{\sigma}_{n}^{y}\hat{\sigma}_{n+1}^{y}\right\}
	-\sum_{n}\frac{\delta_{1}\cos(\omega t)}{4}\left(\hat{\sigma}_{n}^{x}\hat{\sigma}_{n+1}^{y}+\hat{\sigma}_{n}^{y}\hat{\sigma}_{n+1}^{x}\right)-\frac{\delta_{2}}{2}\sum_{n}\hat{\sigma}_{n}^{z}.\label{eq:SCHt}
\end{equation}
\end{widetext}
Here $\delta_{1}$ controls the driving amplitude and $\omega$ is
the driving frequency. $\delta_{2}$ describes the amplitude of magnetic
field along $z$-axis. $\hat{\sigma}_{n}^{x,y,z}$ are Pauli matrices
of quantum spin-$1/2$ variables on the $n$th lattice site. In a
former work \cite{YangPRB2019}, this model has been experimentally
realized to study Floquet dynamical quantum phase transitions and
establish their relations with Floquet topological phases. Performing
Jordan-Wigner and Fourier transformations sequentially under the periodic
boundary condition (PBC), we can express $\hat{H}(t)$ in the Nambu
spinor representation as $\hat{H}(t)=\sum_{k}\hat{\Psi}_{k}^{\dagger}H(k,t)\hat{\Psi}_{k}$,
where $k\in[-\pi,\pi)$ is the quasimomentum and \cite{YangPRB2019}
\begin{equation}
	H(k,t)=d_{x}(k)[\cos(\omega t)\sigma_{x}+\sin(\omega t)\sigma_{y}]+d_{z}(k)\sigma_{z}.\label{eq:SCHkt}
\end{equation}
Here $\sigma_{x,y,z}$ are Pauli matrices in their usual representations
and 
\begin{equation}
	d_{x}(k)=\frac{\delta_{1}\sin k}{2},\qquad d_{z}(k)=\frac{\delta_{1}\cos k+\delta_{2}}{2}.\label{eq:SCdxdz}
\end{equation}
Applying a rotation $|\psi(k,t)\rangle=U_{R}(t)|\varphi(k,t)\rangle$
to the evolving state in the Schr\"odinger equation $i\partial_{t}|\psi(k,t)\rangle=H(k,t)|\psi(k,t)\rangle$,
we could describe the dynamics of rotated state $|\varphi(k,t)\rangle$
by the equation $i\partial_{t}|\varphi(k,t)\rangle=H(k)|\varphi(k,t)\rangle$
with a time-independent Floquet-Bloch effective Hamiltonian $H(k)$,
where $U_{R}(t)={\rm diag}(1,e^{i\omega t})$ and 
\begin{equation}
	H(k)=h_{0}\sigma_{0}+h_{x}(k)\sigma_{x}+h_{z}(k)\sigma_{z},\label{eq:SCHk}
\end{equation}
with $h_{0}=\omega/2$, 
\begin{equation}
	h_{x}(k)=d_{x}(k),\qquad h_{z}(k)=d_{z}(k)-\frac{\omega}{2},\label{eq:SChxhz}
\end{equation}
and $\sigma_{0}$ denotes the two by two identity matrix. Due to the
time-periodicity of $U_{R}(t)=U_{R}(t+T)$ with $T=2\pi/\omega$,
the stroboscopic dynamics of the system is fully governed by $H(k)$,
whose quasienergy bands have the dispersions described by Eq.~(\ref{eq:Esk}) with
\begin{equation}
	E(k)=\sqrt{h_{x}^{2}(k)+h_{z}^{2}(k)}\mod2\pi.\label{eq:SCEk}
\end{equation}
Note in passing that the term $h_0\sigma_0$ in Eq.~(\ref{eq:SCHk}) is generated by the rotating-frame transformation $U_R(t)$. As $h_0\sigma_0$ is proportional to the identity and independent of $k$, it does not affect the geometric and topological properties of Floquet states in our system. In the meantime, if we go back to the original time frame and consider a one-period evolution starting at $t=0$, the Floquet operator of the system becomes $U(k)=e^{-ih_0 T\sigma_0}e^{-i(h_x\sigma_x+h_z\sigma_z)T}=-e^{-i(h_x\sigma_x+h_z\sigma_z)T}$, which possesses the chiral symmetry $\sigma_y$ as $\sigma_yU(k)\sigma_y=U^{\dagger}(k)$ \cite{YangPRB2019}. We also notice that in the quasienergy dispersion $E(k)$, $h_z(k)$ depends on the driving frequency $\omega$ due to Eq.~(\ref{eq:SChxhz}), making $E(k)$ obviously different from the energy spectrum of the static model.

\begin{figure}
	\begin{centering}
		\includegraphics[scale=0.485]{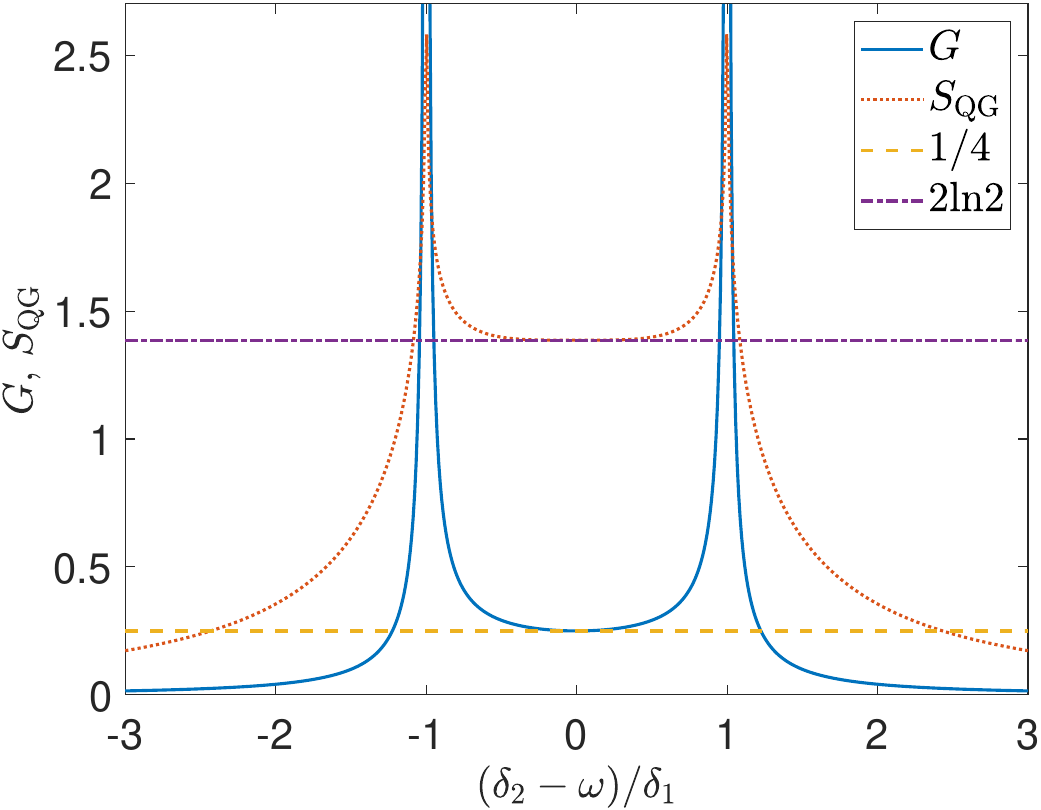}
		\par\end{centering}
	\caption{Integrated QMT $G$ (solid line) and GEE $S_{{\rm QG}}$ (dotted line)
		of the harmonically driven spin chain. The numbers of unit cells and
		filled single-particle states are $L=N=1000$ (half-filling). The
		subsystem size is $L_{A}=500$ (equal bi-partition). The horizontal
		dashed and dash-dotted lines highlight the values of $G$ and $S_{{\rm QG}}$
		in the topological limit $\omega=\delta_{2}$,
		respectively. \label{fig:SCQMT}}
\end{figure}

\begin{figure*}
	\begin{centering}
		\includegraphics[scale=0.35]{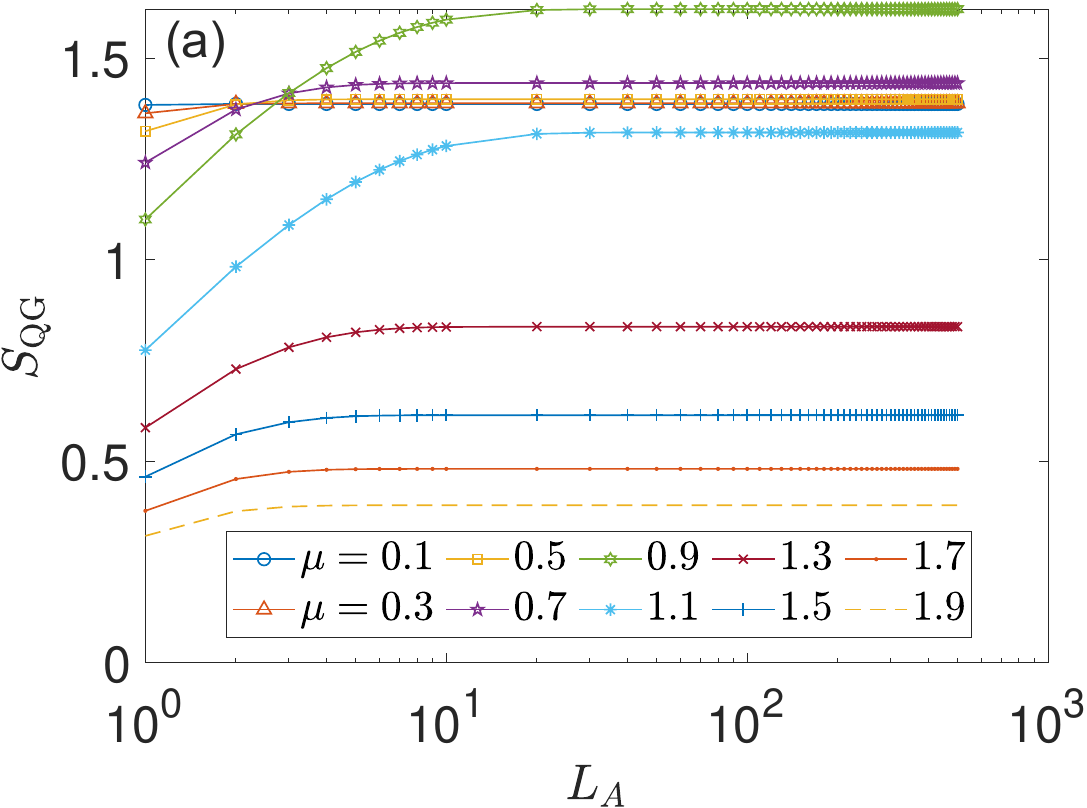}$\,\,\,$\includegraphics[scale=0.35]{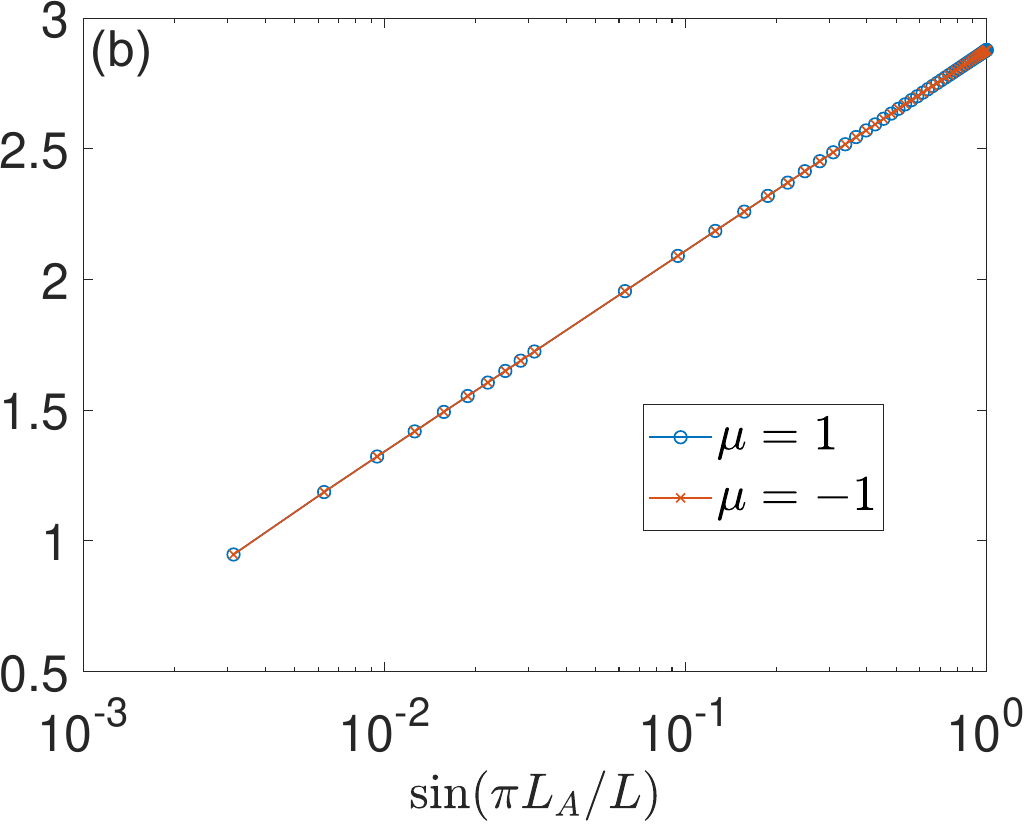}
		\par\end{centering}
	\begin{centering}
		\includegraphics[scale=0.35]{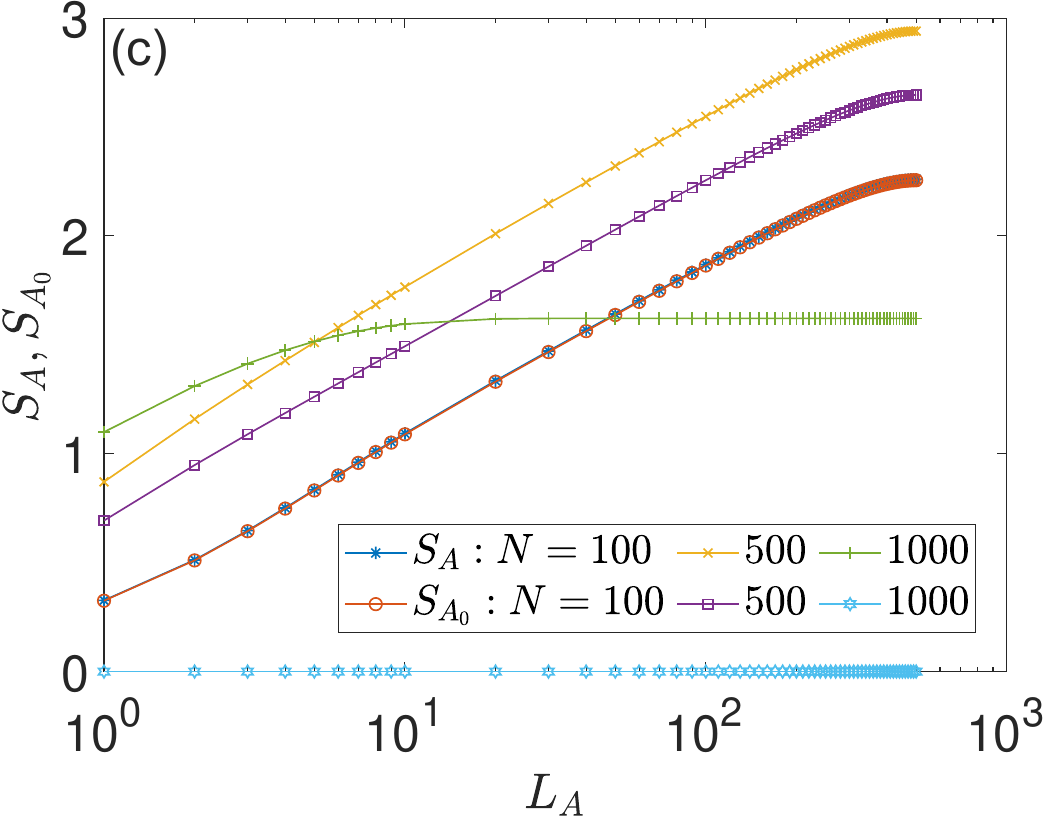}$\,\,$\includegraphics[scale=0.35]{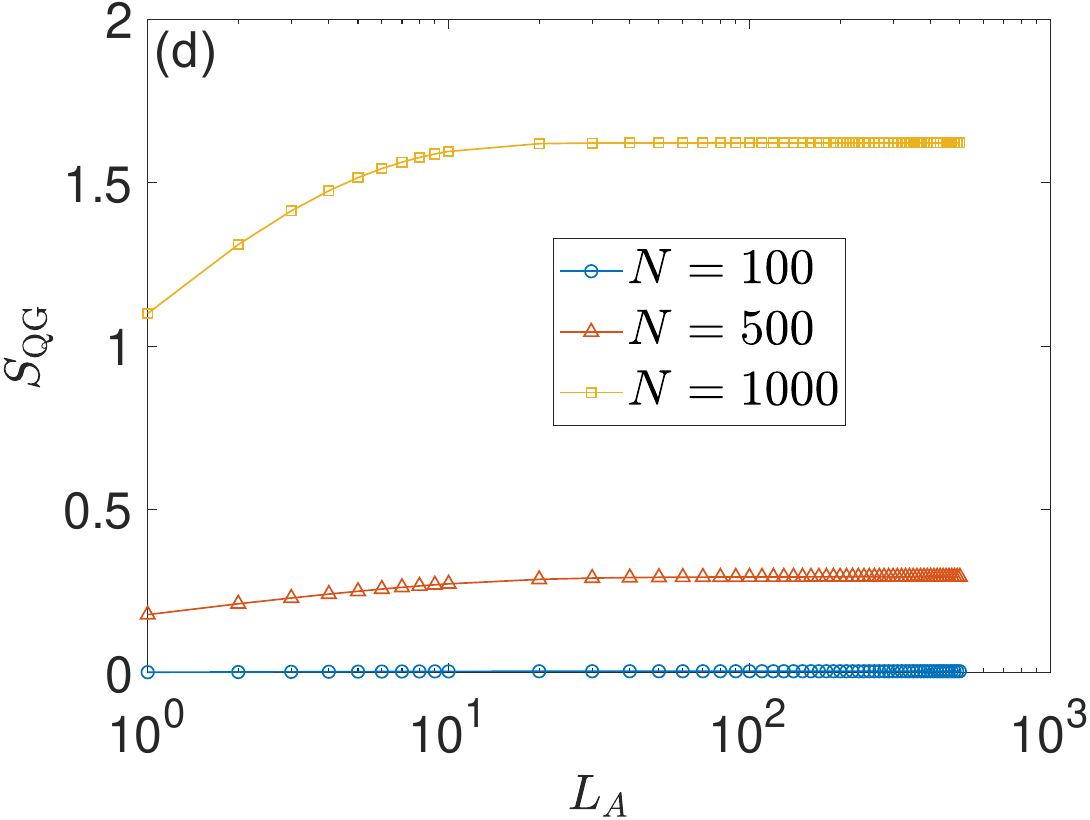}
		\par\end{centering}
	\caption{EE of the harmonically driven spin chain vs the subsystem size $L_{A}$
		under PBC. (a) GEE at half-filling ($N=L$) vs $L_{A}$, with different
		values of $\mu=(\delta_{2}-\omega)/\delta_{1}$ for different curves
		as shown in the figure legend. (b) GEE at half-filling ($N=L$) vs
		$L_{A}$ for $\mu=\pm1$. (c) The total and non-geometric parts of
		EE, $S_{A}$ and $S_{A_{0}}$ vs $L_{A}$ at different filling fractions
		$N/L$, with $L=1000$ and $\mu=0.9$. (d) GEE vs $L_{A}$ at different
		filling fractions $N/L$, with $L=1000$ and $\mu=0.9$. The numbers
		of filled Floquet single-particle states $N$ for different curves
		are shown in the legends of (c) and (d). \label{fig:SCEE}}
\end{figure*}

We can now obtain the QMT and GEE of the system following the Appendices
\ref{sec:Ovlp}--\ref{sec:GEE}. Plugging Eqs.~(\ref{eq:SCdxdz}),
(\ref{eq:SChxhz}) and (\ref{eq:SCEk}) into Eq.~(\ref{eq:gkkZX}),
we find the QMT as
\begin{equation}
	g_{kk}=g_{kk}^{\pm}=\frac{(1+\mu\cos k)^{2}}{4(1+\mu^{2}+2\mu\cos k)^{2}}.\label{eq:SCgkk}
\end{equation}
Here we have set the driving amplitude $\delta_{1}$ as the unit of
energy and introduced the shorthand notation $\mu=(\delta_{2}-\omega)/\delta_{1}$.
Since $g_{kk}\geq0$, we can find its integrated contribution $G$
over the whole BZ, i.e.,
\begin{equation}
	G=\int_{-\pi}^{\pi}\frac{dk}{2\pi}g_{kk}=\begin{cases}
		\frac{1}{8(\mu^{2}-1)}, & |\mu|>1,\\
		\frac{\mu^{2}-2}{8(\mu^{2}-1)}, & |\mu|<1.
	\end{cases}\label{eq:SCgkkInt}
\end{equation}
It is clear that the integrated QMT is divergent at $|\mu|=1$, i.e.,
at $|\delta_{2}-\omega|=|\delta_{1}|$, where we have $E=0$ in 
Eq.~(\ref{eq:SCEk}) and the two Floquet bands $\omega/2\pm E$ meet with
each other at the the quasienergy $\omega/2$, which is the edge of
the first quasienergy BZ. Quantum phase transitions unique to Floquet
systems could then happen at $\mu=\pm1$, which are further associated
with transitions between different Floquet topological phases \cite{YangPRB2019}.
The locations of these transition points are controlled by the driving
frequency $\omega$. The divergence of $G$ at $\mu=\pm1$ then offers
clear geometric signatures for the Floquet topological phase transitions
in our system.

Away from the transition points, we find the limiting behaviors of
$G$ as
\begin{equation}
	\lim_{\mu\rightarrow0}G=\frac{1}{4},\qquad\lim_{\mu\rightarrow\infty}G=0.\label{eq:SCG0inf}
\end{equation}
In Ref.~\cite{YangPRB2019}, it was found that the system belongs
to a topologically nontrivial (trivial) phase when $|\mu|<1$ ($|\mu|>1$).
Therefore, the integration of GMT show different limiting behaviors
in the topological and trivial limits of the system, offering another
geometric probe to distinguish these different Floquet phases. Approaching
the transition point, we find
\begin{equation}
	\lim_{\mu\rightarrow\pm1}G\rightarrow\frac{1}{16}|\mu\mp1|^{-\nu},\label{eq:SCGpm1}
\end{equation}
with the critical exponent $\nu=1$. In Fig.~\ref{fig:SCQMT}, we
present the integrated QMT for a typical set of system parameters,
which clearly demonstrates its critical properties around phase transition
points ($\mu=\pm1$) and limiting behaviors in different Floquet topological
phases. 

Next, to obtain the GEE, we combine the Eqs.~(\ref{eq:SCdxdz}), (\ref{eq:SChxhz})
and (\ref{eq:SCEk}) into Eq.~(\ref{eq:OvlpZX}), yielding the overlap
of wave functions in the Floquet band with dispersion $\omega/2-E(k)$,
i.e., 
\begin{equation}
	\langle\psi_{k}|\psi_{k'}\rangle=\frac{h_{x}(k)h_{x}(k')+[E(k)-h_{z}(k)][E(k')-h_{z}(k')]}{2\sqrt{E(k)E(k')[E(k)-h_{z}(k)][E(k')-h_{z}(k')]}},\label{eq:SCOvlp}
\end{equation}
where $k,k'=2\pi\ell/L$ and $\ell=1,...,N$, with $L$ being the
number of unit cells and $N\leq L$ being the number of occupied single-particle
orbitals in the multi-particle Floquet state $|\Psi\rangle=\prod_{k}|\psi_{k}\rangle$
of our system. For example, when $N=L$, the lower Floquet band with
quasienergy dispersion $\omega/2-E(k)$ is uniformly filled by fermions,
with one at each $k$ in the first BZ. Plugging Eq.~(\ref{eq:SCOvlp})
into Eq.~(\ref{eq:OAkkp3}), diagonalizing the overlap matrices $O^{A}$
and $O^{A_{0}}$ for a given subsystem size $L_{A}$, and inserting
their spectrum into Eqs.~(\ref{eq:SA}) and (\ref{eq:SA0}), we finally
arrive at the von Neumann EE and GEE following Eq.~(\ref{eq:SQG}). For any given
subsystem size $L_{A}\leq L$ and particle number $N\leq L$, we could
then analyze the scaling properties of EE and its behaviors around
phase transition points in our Floquet system.

In Fig.~\ref{fig:SCEE}, we present typical scaling behaviors of EE
and GEE vs the size $L_{A}$ of subsystem $A$. First, we notice that
in either the topological ($|\mu|<1$) or trivial ($|\mu|>1$) phase,
the $S_{{\rm QG}}$ will converge to a finite value that is independent
of $L_{A}$ for the system at half-filling {[}Fig.~\ref{fig:SCEE}(a){]}
or other possible filling fractions {[}Figs.~\ref{fig:SCEE}(c) and
\ref{fig:SCEE}(d){]}. Therefore, the GEE tends to follow an area-law
scaling vs the subsystem size at any filling fractions, so along as
the two Floquet bands of the system are well separated by quasienergy
gaps. The quantum geometric origin of GEE thus endows it with certain
robustness to the change of particle numbers in the system regarding
its scaling properties. Second, at the topological transition points
{[}$\mu=\pm1$ in Fig.~\ref{fig:SCEE}(b){]}, we have $S_{{\rm QG}}\propto\ln[\sin(\pi L_{A}/L)]$.
It suggests that the GEE at half-filling scales logarithmically versus
the subsystem size in these situations (with gapless quasienergy bands),
which is expected for 1D critical metallic phases. Third, in all the
considered cases, we find $S_{A_{0}}=0$ when $N=L$. Therefore, the
bipartite EE of our system at half-filling is solely originated from
the quantum geometry of Floquet-Bloch states in $k$-space. This conclusion
should hold in both Floquet and static systems. Finally, the saturation
value of $S_{{\rm QG}}$ increases monotonically when the system approaches
its topological phase transition point from either side of the parameter
space {[}see Figs.~\ref{fig:SCQMT} and \ref{fig:SCEE}(a){]}. As
will be demonstrated below, the scaling laws and critical properties
of GEE found here are generic and not restricted to the driven spin
chain model considered in this section. Moreover, richer patterns
in QMT and GEE could be identified when the system possesses multiple
Floquet topological phases and phase transitions, as will be considered
in the following sections.

\section{On-resonance double kicked rotor: QMT and GEE\label{sec:ORDKR}}

In the last section, the driven spin chain we introduced owns two
Floquet phases with distinct topological properties. In this section,
we investigate the QMT and GEE of a 1D Floquet insulator with richer
topological phases and transitions \cite{HoPRB2014,LWZPRA2018,LWZPRA2019}. Following
Ref.~\cite{HoPRB2014}, we consider the lattice version of an on-resonance
double kicked rotor (ORDKR), which forms a paradigmatic platform in
the study of dynamical localization, quantum chaos and Floquet topological
matter \cite{HoPRL2012}. The lattice Hamiltonian of such an ORDKR
takes the form \cite{HoPRB2014}
\begin{equation}
	\hat{H}(t)=V(t)\sum_{n}n^{2}\hat{c}_{n}^{\dagger}\hat{c}_{n}+\frac{1}{2}\sum_{n}\left[J(t)\hat{c}_{n}^{\dagger}\hat{c}_{n+1}+{\rm H.c.}\right].\label{eq:DKRHt}
\end{equation}
Here $n\in\mathbb{Z}$ is the lattice index and $\hat{c}_{n}^{\dagger}$
creates a fermion on the lattice site $n$. The onsite potential $V(t)$
and nearest-neighbor hopping amplitude $J(t)$ have the expressions
\begin{equation}
	[V(t),J(t)]=\begin{cases}
		(0,iK_{1}) & t\in[\ell T,\ell T+T/4)\\
		(V,0) & t\in[\ell T+T/4,\ell T+T/2)\\
		(0,K_{2}) & t\in[\ell T+T/2,\ell T+3T/4)\\
		(-V,0) & t\in[\ell T+3T/4,\ell T+T)
	\end{cases},\label{eq:DKRVJ}
\end{equation}
where $K_{1},K_{2},V\in\mathbb{R}$, $\ell\in\mathbb{Z}$ and $T$
is the driving period. In the following calculations, we choose $4\hbar/T$
as the unit of energy and set $V=\pi/2$ in order to obtain a two-band
model. It is clear that the $\hat{H}(t)$ in Eq.~(\ref{eq:DKRHt})
does not have any spatial periodicity. However, the Floquet operator
of the system that governs its evolution over a complete driving period
(e.g., from $t=5T/8$ to $T+5T/8$) takes the form
\begin{alignat}{1}
	\hat{U}= & e^{-\frac{i}{4}\sum_{n}K_{2}(\hat{c}_{n}^{\dagger}\hat{c}_{n+1}+{\rm H.c.})}\nonumber \\
	\times & e^{-\frac{i}{2}\pi\sum_{n}n^{2}\hat{c}_{n}^{\dagger}\hat{c}_{n}}e^{-\frac{i}{2}\sum_{n}iK_{1}(\hat{c}_{n}^{\dagger}\hat{c}_{n+1}-{\rm H.c.})}\nonumber \\
	\times & e^{\frac{i}{2}\pi\sum_{n}n^{2}\hat{c}_{n}^{\dagger}\hat{c}_{n}}e^{-\frac{i}{4}\sum_{n}K_{2}(\hat{c}_{n}^{\dagger}\hat{c}_{n+1}+{\rm H.c.})},\label{eq:DKRU}
\end{alignat}
which has the spatial periodicity under the translation over two lattice
sites ($n\rightarrow n+2$). Performing the Fourier transformation
from position to momentum representations, we can express the Floquet
operator of the system as $\hat{U}=\sum_{k}\hat{\Psi}_{k}^{\dagger}U(k)\hat{\Psi}_{k}$,
where $\hat{\Psi}_{k}^{\dagger}=(\hat{c}_{k,a}^{\dagger},\hat{c}_{k,b}^{\dagger})$
collects creation operators on odd and even sites within each unit
cell and $k$ is the quasimomentum. The Floquet matrix $U(k)$ is
given by \cite{HoPRB2014}
\begin{alignat}{1}
	U(k)= & e^{-\frac{i}{2}{\cal K}_{2}[\cos(k/2)\sigma_{x}+\sin(k/2)\sigma_{y}]}\nonumber \\
	\times & e^{-i{\cal K}_{1}[\sin(k/2)\sigma_{x}-\cos(k/2)\sigma_{y}]}\nonumber \\
	\times & e^{-\frac{i}{2}{\cal K}_{2}[\cos(k/2)\sigma_{x}+\sin(k/2)\sigma_{y}]},\label{eq:DKRUk}
\end{alignat}
where
\begin{equation}
	{\cal K}_{1}\equiv K_{1}\sin(k/2),\qquad{\cal K}_{2}\equiv K_{2}\cos(k/2),\label{eq:DKRK1K2}
\end{equation}
and $\sigma_{x,y}$ are Pauli matrices acting on sublattice degrees
of freedom. Applying the Taylor expansion to each exponential term
in $U(k)$, we find
\begin{equation}
	U(k)=\cos{\cal K}_{1}\cos{\cal K}_{2}-i(h_{x}\sigma_{x}+h_{y}\sigma_{y}),\label{eq:DKRUk2}
\end{equation}
where 
\begin{alignat}{1}
	h_{x}= & \cos\frac{k}{2}\cos{\cal K}_{1}\sin{\cal K}_{2}+\sin\frac{k}{2}\sin{\cal K}_{1},\nonumber \\
	h_{y}= & \sin\frac{k}{2}\cos{\cal K}_{1}\sin{\cal K}_{2}-\cos\frac{k}{2}\sin{\cal K}_{1}.\label{eq:DKRhxhy}
\end{alignat}

\begin{figure}
	\begin{centering}
		\includegraphics[scale=0.49]{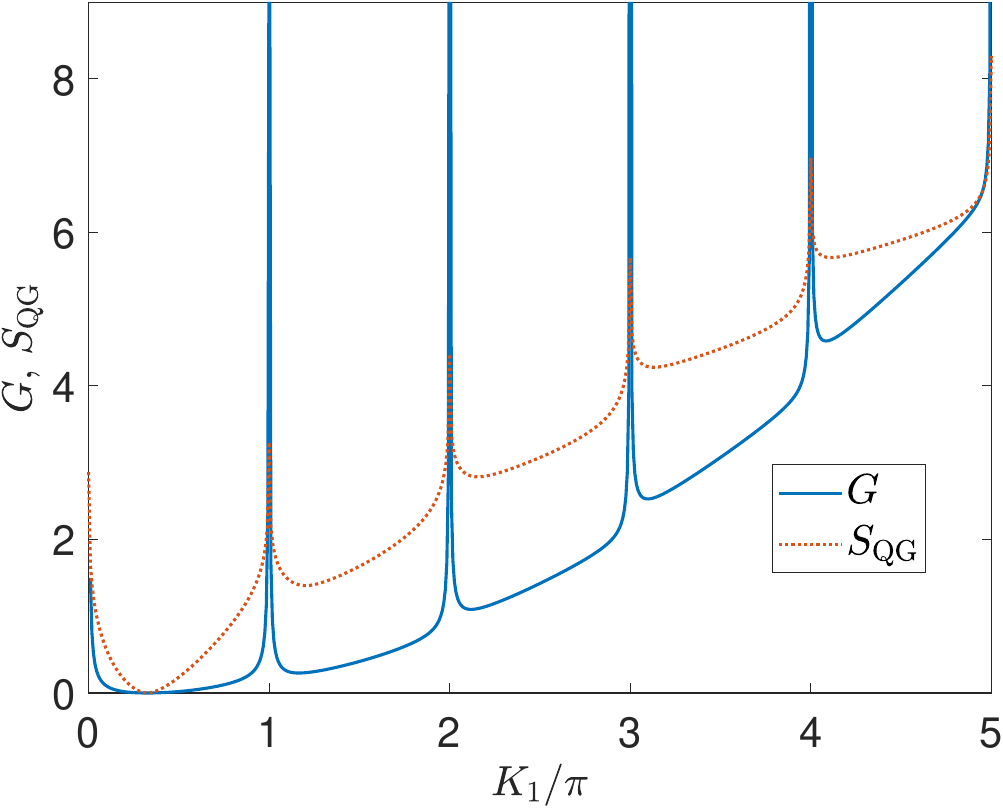}
		\par\end{centering}
	\caption{Integrated QMT $G$ (solid line) and GEE $S_{{\rm QG}}$ (dotted line)
		of the on-resonance double kicked rotor. The kicking strength is $K_{2}=0.5\pi$.
		The numbers of unit cells and filled single-particle states are $L=N=1000$
		(half-filling). The subsystem size is $L_{A}=500$ (equal bi-partition).
		Topological phase transitions happen at $K_{1}=\nu\pi$ {[}Eq. (\ref{eq:DKRPB}){]}
		for $\nu\in\mathbb{Z}$. \label{fig:DKRQMT}}
\end{figure}

\begin{figure*}
	\begin{centering}
		\includegraphics[scale=0.35]{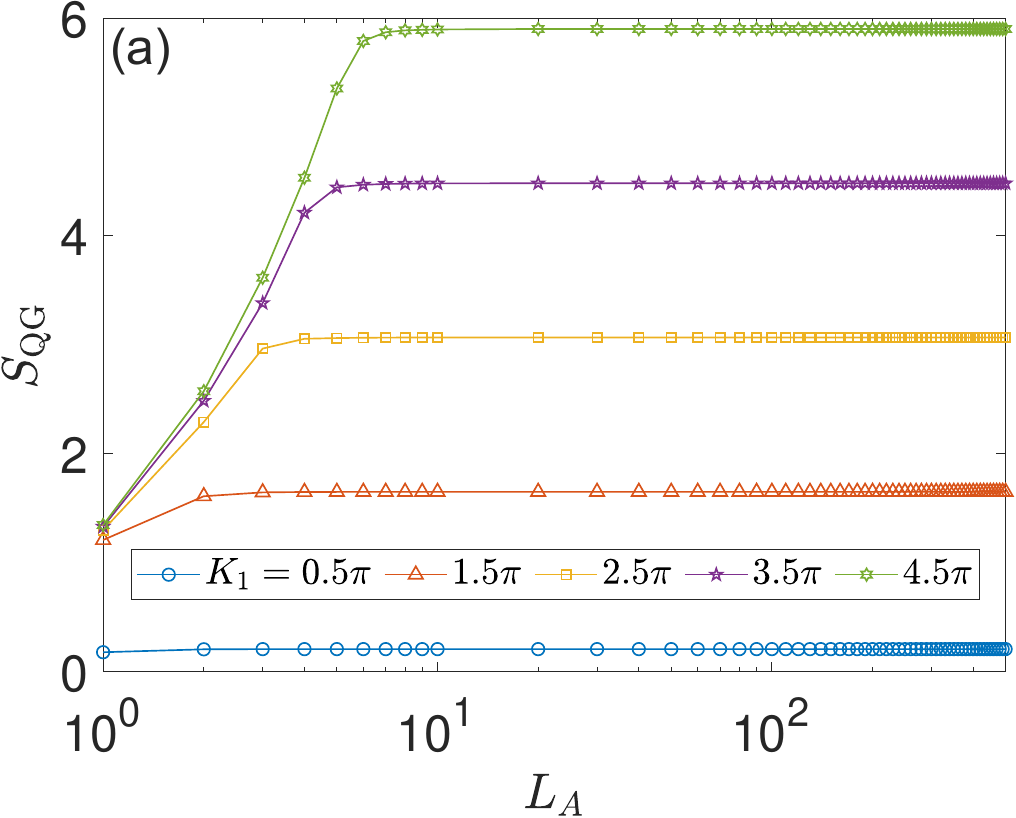}$\,\,\,\,$\includegraphics[scale=0.35]{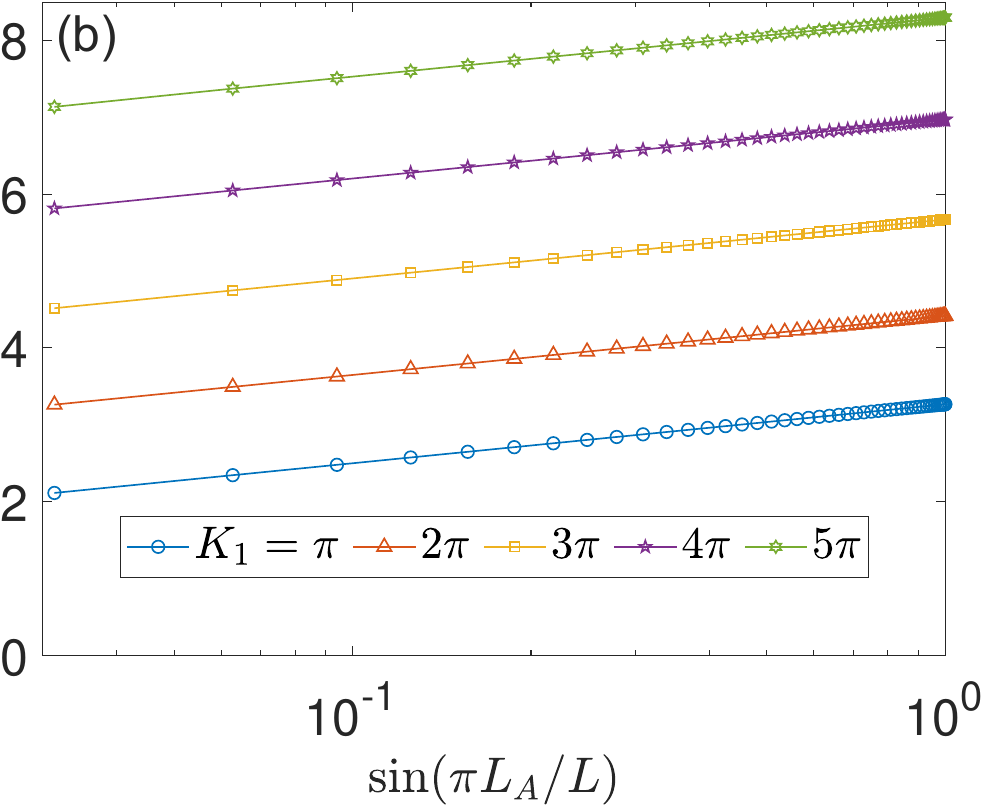}
		\par\end{centering}
	\begin{centering}
		\includegraphics[scale=0.35]{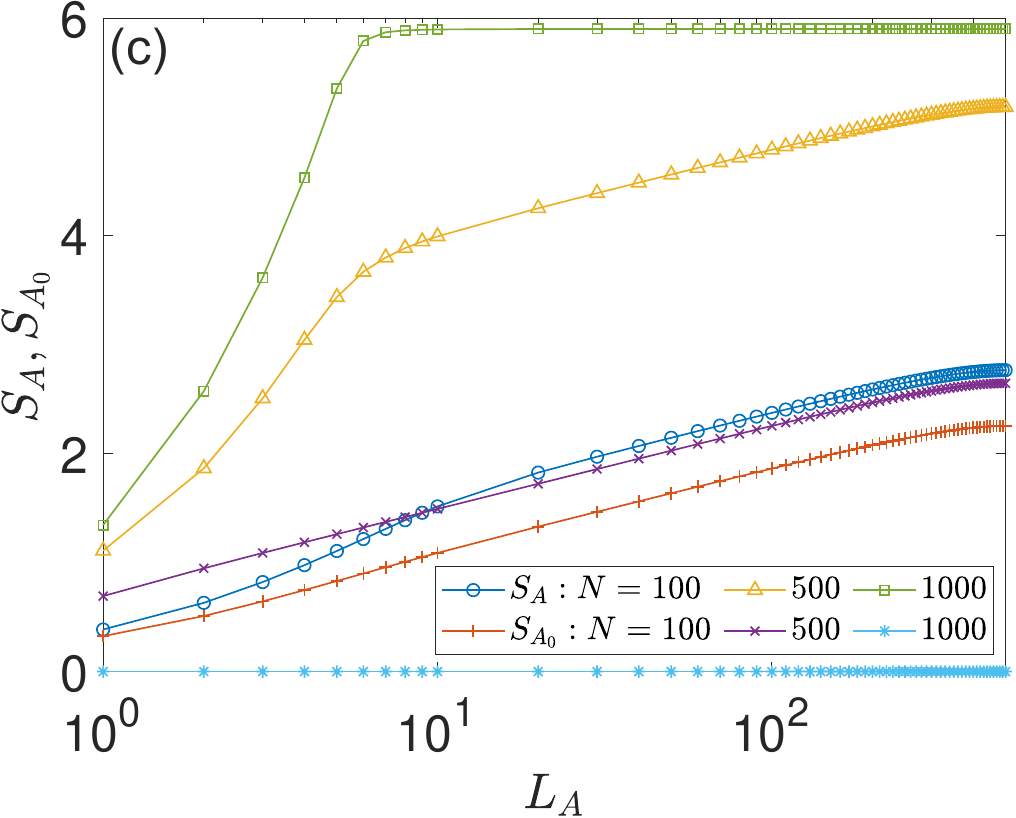}$\,\,$\includegraphics[scale=0.35]{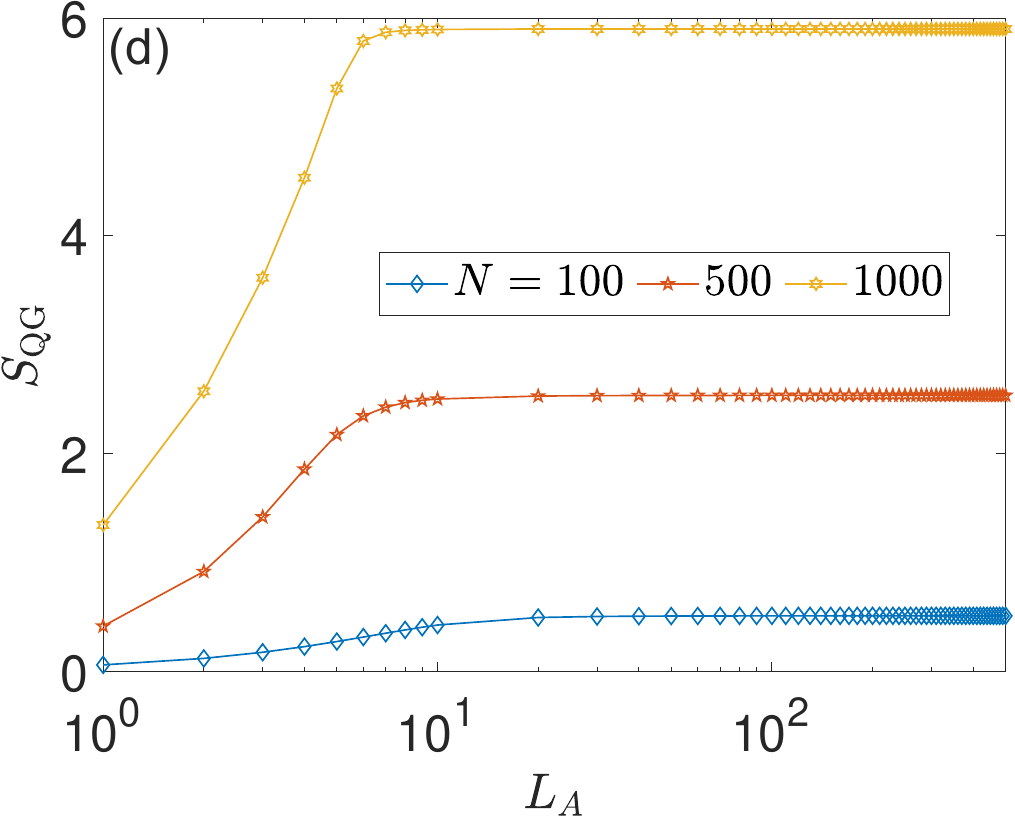}
		\par\end{centering}
	\caption{EE of the on-resonance double kicked rotor vs the subsystem size $L_{A}$
		under PBC, with $K_{2}=0.5\pi$ for all panels. (a) GEE at half-filling
		($N=L$) vs $L_{A}$, with different values of $K_{1}$ for different
		curves as shown in the figure legend. (b) GEE at half-filling ($N=L$)
		vs $L_{A}$ for $K_{1}=\pi,2\pi,3\pi,4\pi$. (c) The total and non-geometric
		parts of EE, $S_{A}$ and $S_{A_{0}}$, vs $L_{A}$ at different filling
		fractions $N/L$, with $L=1000$ and $K_{1}=4.5\pi$. (d) GEE vs $L_{A}$
		at different filling fractions $N/L$, with $L=1000$ and $K_{1}=4.5\pi$.
		The numbers of filled Floquet single-particle states $N$ for different
		curves are shown in the legends of (c) and (d). \label{fig:DKREE}}
\end{figure*}

The quasienergy spectrum of our system thus contains two Floquet bands
under the PBC with dispersion relations $E_{\pm}(k)=\pm\arccos(\cos{\cal K}_{1}\cos{\cal K}_{2})$.
These two bands could touch with each other at the center or boundary
of the first quasienergy BZ $E_{\pm}\in[-\pi,\pi)$, leading to the
gap closing conditions $\cos{\cal K}_{1}\cos{\cal K}_{2}=1$ and $-1$
at the zero and $\pi$ quasienergies, respectively. The boundary curves
between different Floquet topological insulating phases can further
be found as \cite{HoPRB2014}
\begin{equation}
	\frac{\nu^{2}\pi^{2}}{K_{1}^{2}}+\frac{\mu^{2}\pi^{2}}{K_{2}^{2}}=1,\qquad\nu,\mu\in\mathbb{Z}.\label{eq:DKRPB}
\end{equation}
When $K_{1}$ or $K_{2}$ is swept across a phase boundary, the topological
invariants of ORDKR will get quantized changes, which are associated
with the variation of degenerate Floquet zero and $\pi$ edge modes
in the system under the open boundary condition \cite{HoPRB2014}.

The Floquet eigenstates of $U(k)$ are obtained by solving the eigenvalue
equation $U(k)|\psi_{\pm}(k)\rangle=e^{-iE_{\pm}(k)}|\psi_{\pm}(k)\rangle$.
As $U(k)$ and the term $h_{x}\sigma_{x}+h_{y}\sigma_{y}$ in 
Eq.~(\ref{eq:DKRUk2}) commute, they share the same eigenbasis. We could
then focus on the eigenstates of $h_{x}\sigma_{x}+h_{y}\sigma_{y}$
in Eq.~(\ref{eq:DKRUk2}) in order to reveal the quantum geometry
and geometric EE of ORDKR.

To obtain the QMT of ORDKR, we plug Eqs.~(\ref{eq:DKRK1K2}), (\ref{eq:DKRhxhy})
and $E=\sqrt{h_{x}^{2}+h_{y}^{2}}$ into Eq.~(\ref{eq:gkkXY}). Integrating
the resulting $g_{kk}=g_{kk}^{xy}$ over the first BZ yields the integrated
QMT of a filled Floquet quasienergy band, i.e., $G=\int_{-\pi}^{\pi}\frac{dk}{2\pi}g_{kk}$.
In Fig.~\ref{fig:DKRQMT}, we present $G$ vs the kicking strength
$K_{1}$ with $K_{2}=0.5\pi$. According to Eq.~(\ref{eq:DKRPB}),
the quasienergy gap between two Floquet bands of the system closes
when $K_{1}=\nu\pi$ for $\nu\in\mathbb{Z}$ in this case. We observe
that the integrated QMT becomes diverge at these transition points.
This is true for other combinations of system parameters satisfying
Eq.~(\ref{eq:DKRPB}). Notably, the divergent behaviors in $G$ appears
when the Floquet bands touch at either the quasienergy zero (with
$K_{1}=2\nu\pi$) or $\pi$ (with $K_{1}=(2\nu-1)\pi$). Therefore,
the QMT could show non-analytic signatures when the ORDKR undergoes
both normal and anomalous topological transitions between different
Floquet phases, offering a quantum geometric probe to these nonequilibrium
phase transitions. To find the GEE, we first obtain the overlap between
any two Floquet eigenstates in the lower quasienergy band, i.e., ${\cal O}_{-}^{xy}(k,k')$,
according to Eq.~(\ref{eq:OvlpXY}), where $E=\sqrt{h_{x}^{2}+h_{y}^{2}}$
and the $h_{x},h_{y}$ are given by Eq.~(\ref{eq:DKRhxhy}). Replacing
the $\langle\psi_{k}|\psi_{k'}\rangle$ in Eq.~(\ref{eq:OAkkp3})
by ${\cal O}_{-}^{xy}(k,k')$ and following the recipes in the Appendices
\ref{sec:EE}--\ref{sec:GEE}, we can obtain the total EE together
with its geometric and non-geometric parts from the eigenspectrum
of overlap matrix $O^{A}$ in $k$-space. 

The GEE versus $K_{1}$ at half-filling and under equal bipartition
for the ORDKR is shown in Fig.~\ref{fig:DKRQMT}. We find that around
each topological transition point, the GEE shows a cusp, with a discontinuous
derivative vs $K_{1}$ at different sides of $K_{1}=\nu\pi$. Away
from the topological transition points, the amount of GEE increases
gradually with the increase of $K_{1}$. This is related to the raise
of topological edge-state numbers at zero and $\pi$ quasienergies
following the increase of kicking strengths in ORDKR \cite{HoPRB2014}.
These Floquet edge modes yield gradually increased contributions to
GEE across the entanglement cuts when a bipartition is taken in the
bulk. Putting together, thanks to the close integration among quantum
information, geometry and topology, we could employ the GEE as an
additional probe to the phases and transitions in 1D Floquet topological
insulators with large topological invariants.

To further decode the scaling laws of EE, we could first decompose
the system $S$ into two complementary parts as $S=A\cup\overline{A}$.
The behaviors of EE vs the subsystem size $L_{A}$ and the filling
fraction $N/L$ of a Floquet band can then be worked out, with $N$
the total number of particles and $L=L_{A}+L_{\overline{A}}$ the
system size. Results of EE for typical cases of the ORDKR are shown
in Fig.~\ref{fig:DKREE}, which are obtained following the procedure
in Appendices \ref{sec:EE}--\ref{sec:GEE}. First, we notice that
when the system at half-filling resides in gapped Floquet topological
insulator phases {[}the cases in Fig.~\ref{fig:DKREE}(a){]}, the
GEE would always converge to an area-law scaling vs the system size
with the increase of $L_{A}$. In contrast, when the two Floquet bands
meet at the quasienergy zero or $\pi$ {[}the cases in Fig.~\ref{fig:DKREE}(b){]},
the system becomes critical at half-filling and the GEE follows a
log-law scaling vs $L_{A}$. These different scaling laws (area-law
and log-law) of GEE are generic when the ORDKR is prepared in other
gapped and gapless Floquet topological phases, respectively. A further
comparison between the results in Figs.~\ref{fig:DKREE}(c) and \ref{fig:DKREE}(d)
suggests that the bipartite EE of ORDKR is fully quantum geometric
if the many-particle state of the system uniformly fills a Floquet
band. That is, the non-geometric EE $S_{A_{0}}$ vanishes and the
total EE $S_{A}$ becomes equal to $S_{{\rm QG}}$ at half-filling
{[}$N/L=1$ in Figs.~\ref{fig:DKREE}(c) and \ref{fig:DKREE}(d){]}.
This observation clarifies the geometric origin of EE in 1D Floquet
topological insulators, i.e., the bipartite EE of a filled Floquet
band is uniquely determined by the quantum geometry of the populated
Floquet-Bloch eigenstates. Finally, even though both the bipartite
EE and its non-geometric part could vary with the subsystem size $L_{A}$,
the GEE always satisfies an area-law versus the subsystem size. This
is true regardless of the filling fraction ($N/L\in(0,1]$) of the
considered Floquet band, so long as it is gapped from the other bands.
The scaling behavior of GEE may thus be robust to the variations of
Floquet-band populations through certain dynamical processes, which
further highlights its geometric origin.

In comparison with the driven spin chain studied in the last section,
the QMT and GEE found here show quantitatively richer patterns due
to the underlying multiple Floquet topological insulating phases and
transitions within the ORDKR. Meanwhile, the generic scaling and critical
properties of QMT and EE in these two models are coincident. These general
relations will hold also in 1D Floquet topological superconductors,
as will be unveiled in the following section.

\section{Periodically quenched Kitaev chain: QMT and GEE\label{sec:PQKC}}

\begin{figure}
	\begin{centering}
		\includegraphics[scale=0.485]{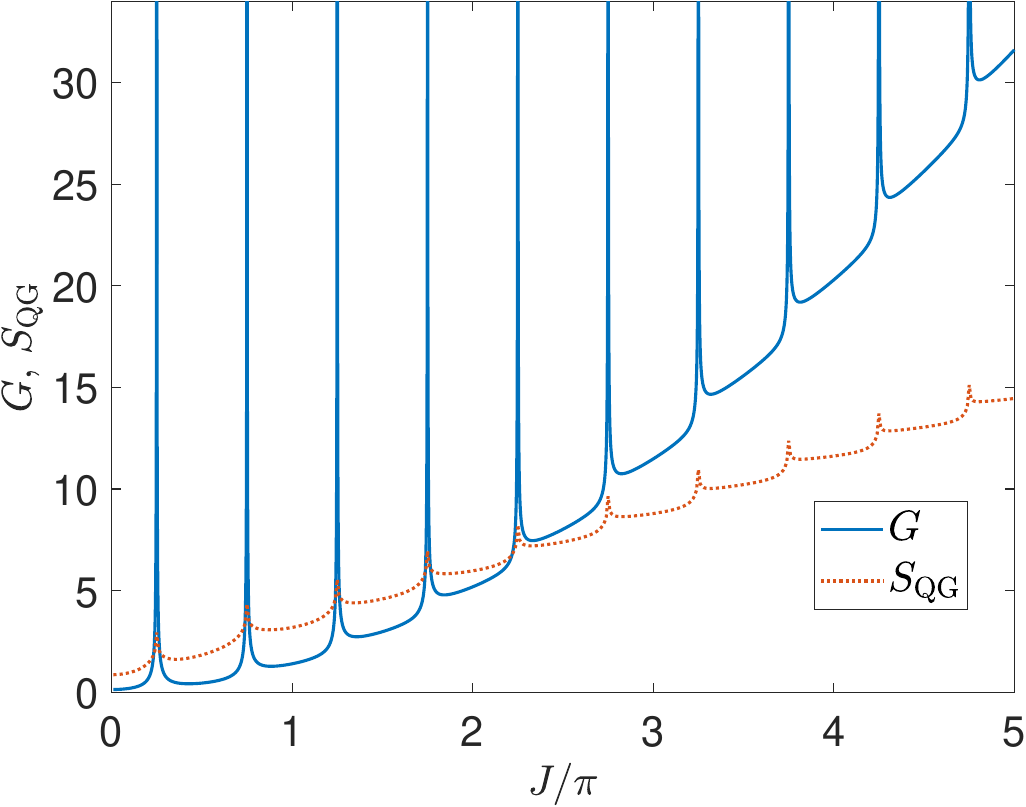}
		\par\end{centering}
	\caption{Integrated QMT $G$ (solid line) and GEE $S_{{\rm QG}}$ (dotted line)
		of the periodically quenched Kitaev chain. Other system parameters
		are $\Delta=\pi/2$ and $\mu=\pi/4$. The numbers of unit cells and
		filled single-particle states are $L=N=1000$ (half-filling). The
		subsystem size is $L_{A}=500$ (equal bi-partition). Topological phase
		transitions happen at $J=\pi/4+\nu\pi/2$ for $\nu\in\mathbb{Z}$
		{[}Eq.~(\ref{eq:KCPB}){]}. \label{fig:KCQMT}}
\end{figure}

\begin{figure*}
	\begin{centering}
		\includegraphics[scale=0.35]{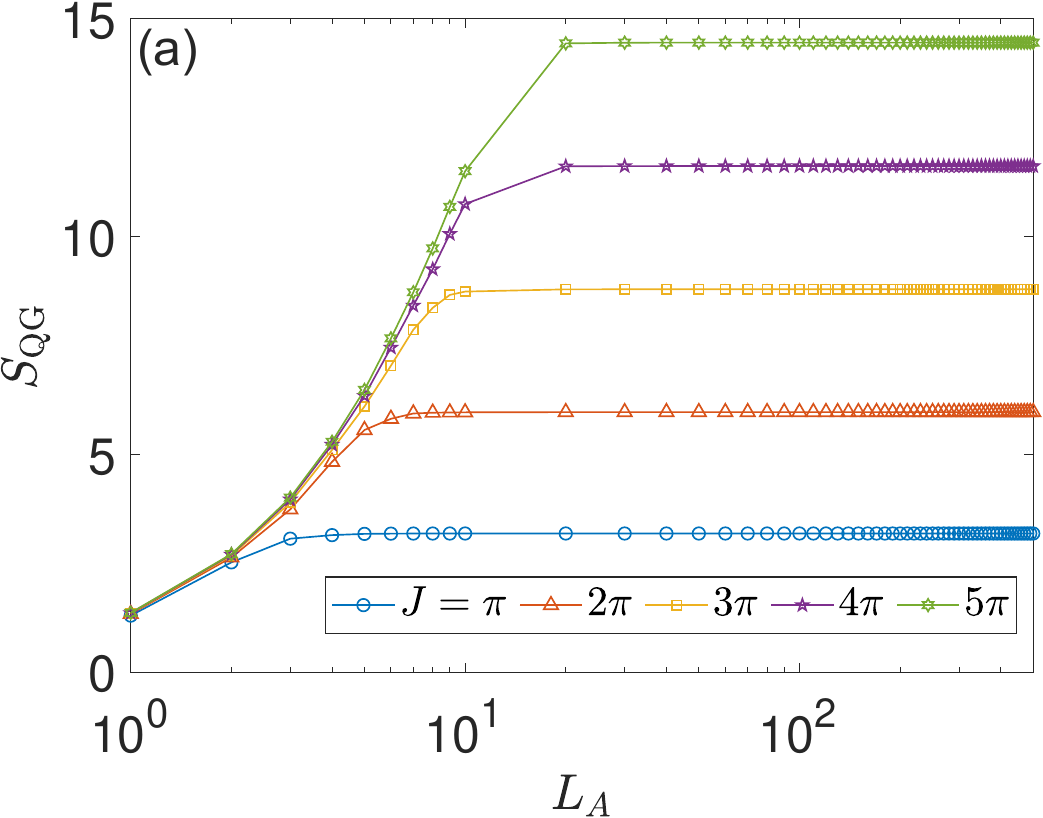}$\,\,\,\,\,$\includegraphics[scale=0.35]{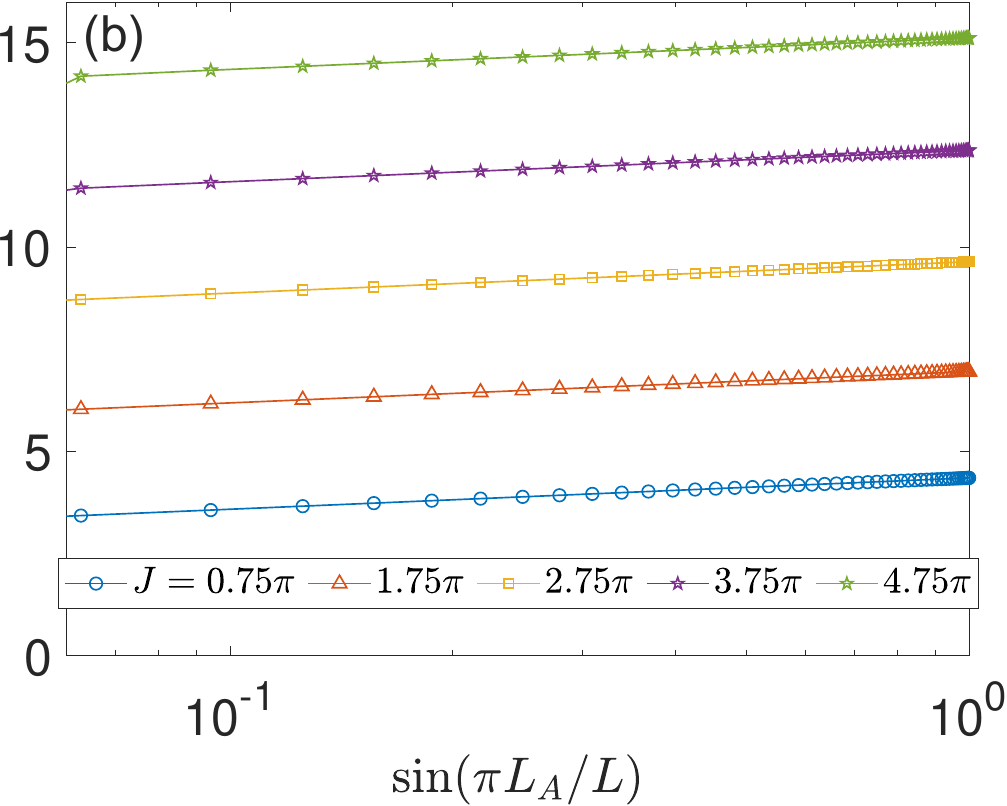}
		\par\end{centering}
	\begin{centering}
		\includegraphics[scale=0.35]{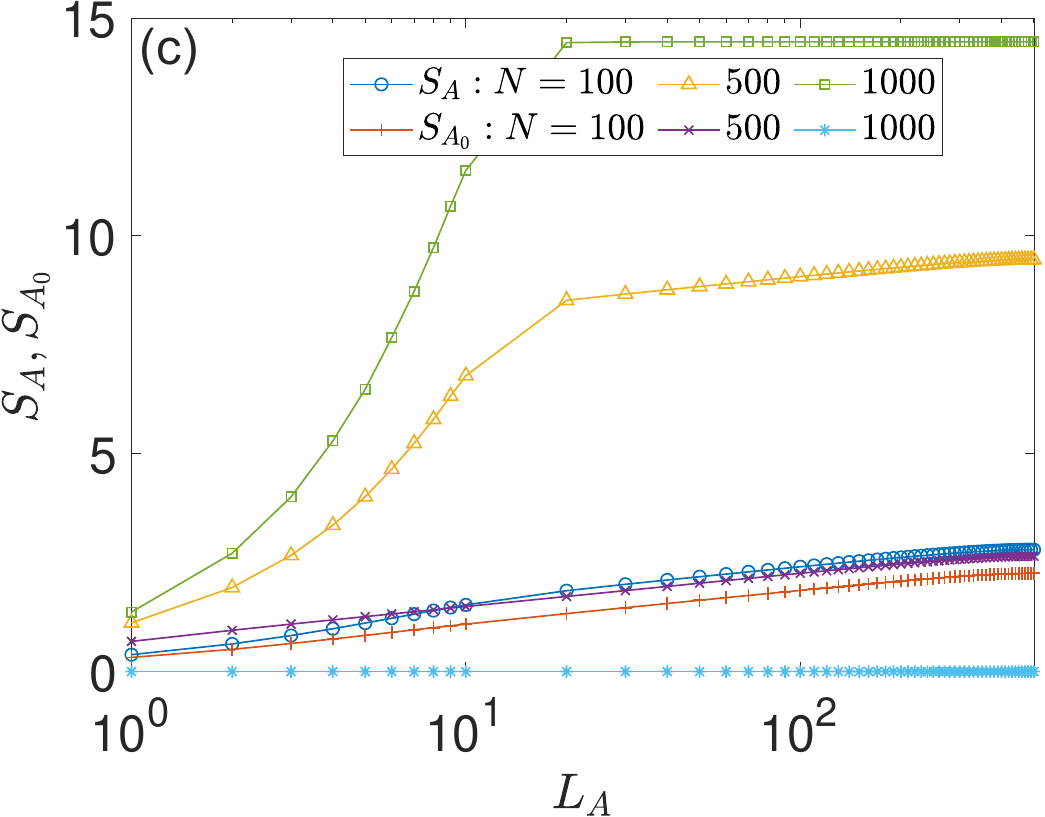}$\,\,$\includegraphics[scale=0.35]{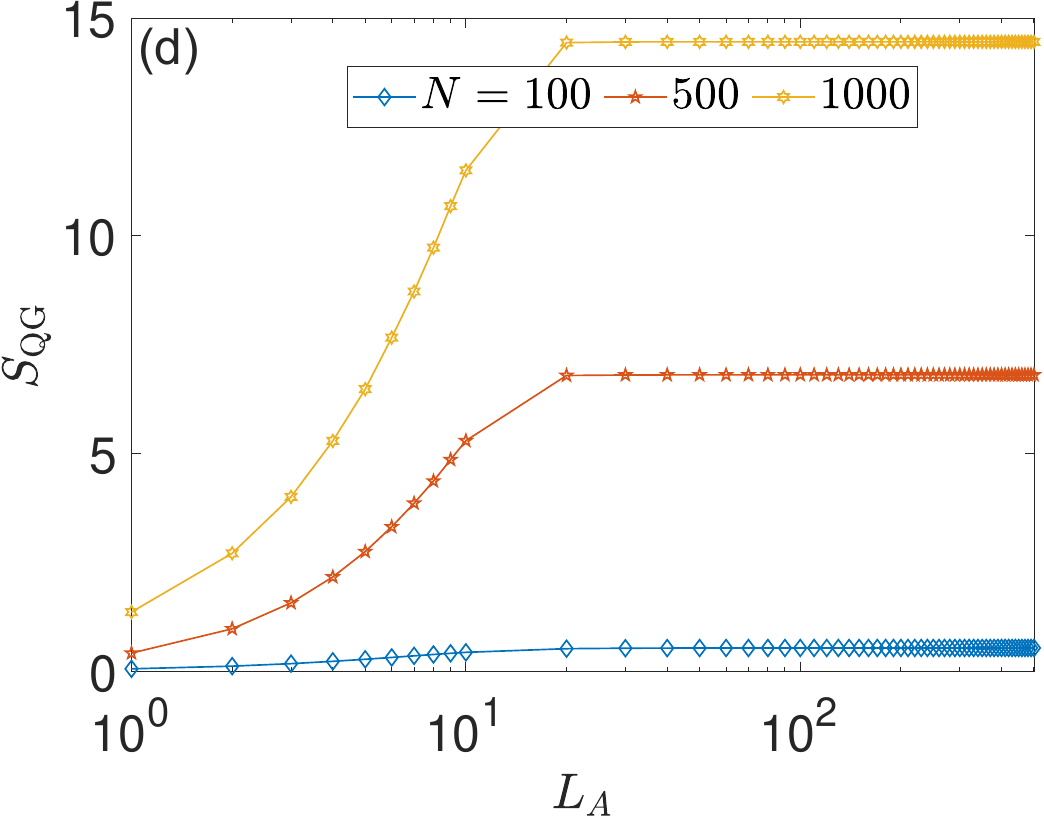}
		\par\end{centering}
	\caption{EE of the periodically quenched Kitaev chain vs the subsystem size
		$L_{A}$ under PBC. Other system parameters are $\Delta=\pi/2$ and
		$\mu=\pi/4$ for all panels. (a) GEE at half-filling ($N=L$) vs $L_{A}$,
		with different values of $J$ for different curves as shown in the
		figure legend. (b) GEE at half-filling ($N=L$) vs $L_{A}$ for $J=0.75\pi+\nu\pi$
		and $\nu=0,1,2,3,4$. (c) The total and non-geometric parts of EE,
		$S_{A}$ and $S_{A_{0}}$ vs $L_{A}$ at different filling fractions
		$N/L$, with $L=1000$ and $J=5\pi$. (d) GEE vs $L_{A}$ at different
		filling fractions $N/L$, with $L=1000$ and $J=5\pi$. The numbers
		of filled Floquet single-particle states $N$ for different curves
		are shown in the legends of (c) and (d). \label{fig:KCEE}}
\end{figure*}

We now consider the quantum geometry and geometric EE of Floquet states
in a periodically quenched Kitaev chain (PQKC) \cite{LWZPRB2020,LWZNJP2023},
whose time-dependent Hamiltonian takes the form
\begin{equation}
	\hat{H}(t)=\begin{cases}
		\hat{H}_{1} & t\in[\ell T,\ell T+T/4)\\
		\hat{H}_{2} & t\in[\ell T+T/4,\ell T+3T/4)\\
		\hat{H}_{1} & t\in[\ell T+3T/4,\ell T+T)
	\end{cases}.\label{eq:KCHt}
\end{equation}
Here $\ell\in\mathbb{Z}$ counts the number of driving period $T$.
The piecewise Hamiltonians
\begin{equation}
	\hat{H}_{1}=\frac{1}{2}\sum_{n}\Delta(\hat{c}_{n}\hat{c}_{n+1}+{\rm H.c.}),\label{eq:KCH1}
\end{equation}
\begin{equation}
	\hat{H}_{2}=\frac{1}{2}\sum_{n}[\mu(\hat{c}_{n}^{\dagger}\hat{c}_{n}-1/2)+J\hat{c}_{n}^{\dagger}\hat{c}_{n+1}+{\rm H.c.}],\label{eq:KCH2}
\end{equation}
where $\hat{c}_{n}^{\dagger}$ creates a fermion on the lattice site
$n\in\mathbb{Z}$. $\Delta$ is the superconducting pairing strength,
$\mu$ is the chemical potential, and $J$ is the nearest-neighbor
hopping amplitude. The Floquet operator of the system, which leads
its evolution over a complete driving period (e.g., from $t=\ell T+0^{-}$
to $t=\ell T+T+0^{-}$) takes the form of $\hat{U}=e^{-i\frac{T}{4\hbar}\hat{H}_{1}}e^{-i\frac{T}{2\hbar}\hat{H}_{2}}e^{-i\frac{T}{4\hbar}\hat{H}_{1}}$.
Under the PBC, we can transform $\hat{U}$ from the position to momentum
representations and express it in terms of the Nambu spinor basis
as $\hat{U}=\sum_{k}\hat{\Xi}_{k}^{\dagger}U(k)\hat{\Xi}_{k}$, where
$k$ is the Bloch quasimomentum and $\hat{\Xi}_{k}^{\dagger}=(\hat{c}_{k}^{\dagger},\hat{c}_{-k})$.
The Floquet operator $U(k)$ in $k$-space reads
\begin{equation}
	U(k)=e^{-\frac{i}{2}d_{y}\sigma_{y}}e^{-id_{z}\sigma_{z}}e^{-\frac{i}{2}d_{y}\sigma_{y}},\label{eq:KCUk}
\end{equation}
where $\sigma_{y,z}$ are Pauli matrices, 
\begin{equation}
	d_{y}=\Delta\sin k,\qquad d_{z}=\mu+J\cos k,\label{eq:KCdydz}
\end{equation}
and we have set $2\hbar/T$ as the unit of energy. Applying the Taylor
expansion to each exponential term of $U(k)$ and recombine the relevant
terms, we find
\begin{equation}
	U(k)=\cos d_{y}\cos d_{z}-i(h_{y}\sigma_{y}+h_{z}\sigma_{z}),\label{eq:KCUk2}
\end{equation}
\begin{equation}
	h_{y}=\sin d_{y}\cos d_{z},\qquad h_{z}=\sin d_{z}.\label{eq:KChyhz}
\end{equation}

The Floquet operator $U(k)$ has two quasienergy bands, whose dispersions
are $E_{\pm}(k)=\pm\arccos(\cos d_{y}\cos d_{z})$. The quasienergy
spectrum of $U(k)$ then becomes gapless when $\cos d_{y}\cos d_{z}=\pm1$,
yielding the borderlines between its different Floquet topological
phases as \cite{LWZNJP2023}
\begin{equation}
	\frac{\kappa^{2}\pi^{2}}{\Delta^{2}}+\frac{(\nu\pi-\mu)^{2}}{J^{2}}=1,\qquad\kappa,\nu\in\mathbb{Z}.\label{eq:KCPB}
\end{equation}
Meanwhile, we notice that the matrix $h(k)\equiv h_{y}\sigma_{y}+h_{z}\sigma_{z}$
in Eq.~(\ref{eq:KCUk2}) commutes with the Floquet operator $U(k)$.
They thus share the same Floquet eigenstates. This allows us to deduce
the QMT and GEE of our PQKC model from the eigenbasis of $h(k)$. 

Plugging Eqs.~(\ref{eq:KCdydz}) and (\ref{eq:KChyhz}) into Eq.~(\ref{eq:gkkYZ}),
we can find the QMT of our system analytically as
\begin{equation}
	g_{kk}=\frac{[J\sin k\sin d_{y}+(\Delta/2)\cos k\cos d_{y}\sin(2d_{z})]^{2}}{4(\sin^{2}d_{y}\cos^{2}d_{z}+\sin^{2}d_{z})^{2}}.\label{eq:KCgkk}
\end{equation}
The integrated contribution of $g_{kk}$ over the first BZ, i.e.,
$G=\int_{-\pi}^{\pi}\frac{dk}{2\pi}g_{kk}$ can be further obtained
numerically. In Fig.~\ref{fig:KCQMT}, we present $G$ vs the hopping
amplitude $J$ for a typical set of system parameters. We find that
the QMT becomes divergent at every topological phase transition point
in the system, where the gap between $E_{\pm}(k)$ closes at the quasienergy
zero (with $J=2\nu\pi\pm\pi/4$ and $\nu\in\mathbb{Z}$) or $\pi$
(with $J=(2\nu-1)\pi\pm\pi/4$ and $\nu\in\mathbb{Z}$). Therefore,
both the normal and anomalous topological transitions between different
Floquet superconducting phases could yield non-analytic signatures
in the QMT of a filled Floquet band. The latter could thus supply
a quantum geometric probe to the phase transitions in Floquet topological
superconductors. 

To find out the GEE, we could first insert Eq.~(\ref{eq:KChyhz})
and $E(k)=\sqrt{h_{y}^{2}+h_{z}^{2}}$ into Eq.~(\ref{eq:OvlpYZ}),
yielding the overlap ${\cal O}_{-}^{yz}(k,k')$ between eigenstates
in the lower Floquet band with quasienergy dispersion $E_{-}(k)$.
Replacing the term $\langle\psi_{k}|\psi_{k'}\rangle$ with ${\cal O}_{-}^{yz}(k,k')$
in Eq.~(\ref{eq:OAkkp3}) gives us the wave function overlap $O_{k,k'}^{A}$,
from which the total, non-geometric and geometric parts of EE could
be deduced following the Appendices \ref{sec:EEvsOA} and \ref{sec:GEE}.
The dotted line in Fig.~\ref{fig:KCQMT} shows the change of GEE with
respect to $J$ for the PQKC at half-filling and under equal bi-partition.
We find that the amount of $S_{{\rm QG}}$ gradually raises with the
increase of $J$, and at each topological transition point it exhibits
a cusp structure. The former is related to the fact that the topological
invariants of PQKC could increase monotonically following a sequence
of topological transitions triggered by the increase of $J$, generating
more and more Floquet Majorana edge modes at zero and $\pi$ quasienergies
that could contribute to EE when a bi-partition is taken in the bulk
\cite{LWZNJP2023}. This is in stark contrast to the results shown
in Fig.~\ref{fig:SCQMT}, where the system only possesses two different
topological phases. The observed cusps in $S_{{\rm QG}}$ further
imply that one can use GEE as a detector for the topological transitions
between distinct Floquet superconducting phases from an integrated
view of quantum geometry and information.

To further decode the scaling properties of EE in our PQKC, we present
in Fig.~\ref{fig:KCEE} the total, geometric and non-geometric parts
of EE vs the subsystem size $L_{A}$ for some typical cases, both
away from and at topological transition points. The calculations of
EE also follow the Appendices \ref{sec:EEvsOA} and \ref{sec:GEE}. In
Fig.~\ref{fig:KCEE}(a), we observe that in gapped Floquet superconducting
phases, the $S_{{\rm QG}}$ will finally converge to a value for each
case that is increasing with $J$ but independent of $L_{A}$. Therefore,
the $S_{{\rm QG}}$ at half-filling follows an area-law scaling vs
the system size. A further comparison between Figs.~\ref{fig:KCEE}(c)
and \ref{fig:KCEE}(d) suggests that this area-law scaling behavior
is independent of the filling fraction $N/L$ of the considered Floquet
band. Therefore, in topological phases with gapped Floquet bands,
the GEE of PQKC follows an area-law regardless of the filling fraction
of the band. This is consistent with our results in Secs.~\ref{sec:HDSC}
and \ref{sec:ORDKR} for other Floquet models. Next, we notice that
the GEE follows sub-volume log-law scalings at the critical points
between different Floquet superconducting phases, as shown in 
Fig.~\ref{fig:KCEE}(b). This is true regardless of whether the Floquet
bands close their respective gaps at the center ($E=0$) or boundary
($E=\pi$) of the first quasienergy BZ. Referring to the results in
Figs.~\ref{fig:SCEE}(b) and \ref{fig:DKREE}(b), we find that the
log-law scalings of GEE at topological transition points of 1D Floquet
phases can be satisfied in general. Third, away from the half-filling,
both the total and non-geometric parts of EE are sensitive to the
change of subsystem size $L_{A}$, as show in Fig.~\ref{fig:KCEE}(c).
However, at half-filling, the non-geometric EE $S_{A_{0}}$ vanishes,
and the total EE becomes fully quantum geometric, as observable from
the curves with $N=L$ in Figs.~\ref{fig:KCEE}(c) and \ref{fig:KCEE}(d).
This is also coincident with our results for the harmonically driven
spin chain and double kicked rotor in the last two sections. Therefore,
the bipartite EE for states filling a gapped quasienergy band tends
out to be solely of quantum geometric origin and satisfies an area-law
scaling in 1D Floquet systems. Physical properties associated with
such GEE should then be robust to certain dynamical variations of
the system, yielding auxiliary probes to the topological phase transitions
in 1D Floquet superconducting systems.

\section{Summary and discussion\label{sec:Sum}}

In this work, we revealed the quantum geometry and geometric entanglement
entropy of typical Floquet topological phases in 1D systems. Based
on the detailed calculations of quantum geometric tensors, entanglement
entropy and their scaling behaviors for periodically driven spin chains,
Floquet topological insulators and superconductors, we could arrive
at the following general conclusions.

First, for a uniformly filled Floquet quasienergy band, the quantum metric
tensor integrated over the occupied Floquet-state manifold becomes
divergent at the transition points between different Floquet topological
phases. The quantum geometry of Floquet-Bloch states could thus offer
an efficient means to probe topological phase transitions in Floquet
systems. Second, regardless of the filling fractions of a gapped Floquet-Bloch
band, the geometric EE as considered in this study always follow an
area-law scaling vs the system size. This observation implies certain
levels of robustness of the geometric EE to the variation of Floquet
band populations, which may find applications in the characterization
of quasienergy bands in particle-number non-conserved (or open-system)
situations. Third, for a Floquet band at unit filling, the bipartite
EE becomes purely quantum-geometric. The EE of Floquet systems at
half-filling, as considered in previous studies \cite{LWZPRR2022},
could thus be viewed as geometric EE, which might be insensitive to
the changes of some dynamical details of the system. Finally, the
bipartite geometric EE of a filled Floquet-Bloch band shows a critical
log-law scaling versus the system size at each topological phase transition
point. Close to the transition point, the geometric EE further exhibits
the shape of a cusp versus the transition-driven parameter of the system.
These observations suggest that the geometric EE could provide us
with an efficient probe to identify Floquet topological phase transitions
from a hybrid perspective of quantum geometry and quantum information.

As an additional comment, for all the models we considered, the two Floquet bands can be separated by two gaps at both the quasienergies zero and $\pi$, instead of a single gap around zero energy in non-driven two-band models. Both the zero and $\pi$ Floquet gaps can be topologically nontrivial and admit degenerate topological edge modes under the open boundary condition \cite{HoPRB2014,LWZNJP2023}. The two Floquet bands could further touch with each other at either the quasienergy zero or $\pi$, causing two possible avenues of topological phase transitions. When the Floquet bands meet and re-separate at the quasienergy $\pi$, an anomalous phase transition that cannot manifest in a static two-band system could happen, whose signatures in quantum geometry and GEE are unique to Floquet systems and are characterized in detailed through our model studies.

The conclusions as mentioned above are expected to be generic and
not restricted to the models considered in this work. The verification
(and possible extension) of these results for Floquet systems in other
symmetry classes and in higher spatial dimensions deserves further
considerations. The properties of Floquet quantum geometry, geometric
EE and their robustness against more complicated effects such as disorders
and interactions constitute interesting directions of future research.
Besides, the experimental detection of quantum metric tensor and geometric
EE of Floquet states could be within reach in various quantum simulators
like nitrogen-vacancy center in diamonds \cite{QGTExpNV1,QGTExpNV2},
superconducting qubits \cite{QGTExpSC1,QGTExpSC2} and ultracold atoms
\cite{QGTExpCA1,QGTExpCA2,QGTExpCA3}.

\begin{acknowledgments}
	This work is supported by the Fundamental Research Funds for the Central Universities (Grant No.~202364008), the National Natural Science Foundation of China (Grants No.~12275260, No.~12047503 and No.~11905211),  and the Young Talents Project of Ocean University of China.
\end{acknowledgments}

\appendix

\section{Wavefunction overlap of 1D, two-band Floquet effective Hamiltonians\label{sec:Ovlp}}

In this Appendix, we compute the overlap between eigenstates within
a single Floquet-Bloch band for 1D lattice models. Under
time-periodic drivings and spatial periodic boundary conditions (PBCs),
the Hamiltonian of such a lattice model takes the form $\hat{H}(t)=\sum_{k}|k\rangle H(k,t)\langle k|$,
where $k\in[-\pi,\pi)$ is the quasimomentum. The Floquet operator
of the system then reads $\hat{U}=\sum_{k}|k\rangle U(k)\langle k|$,
where $U(k)\equiv\hat{\mathsf{T}}e^{-i\int_{t}^{t+T}H(k,t')dt'}$
with $T$ being the driving period. Formally, one can write $U(k)$
as $U(k)=e^{-iH(k)}$, with the effective Floquet-Bloch Hamiltonian
$H(k)$ defined by $H(k)=i\ln U(k)$. For a generic 1D lattice with
two internal degrees of freedom (spins, sublattices, etc.) in each
unit cell, one can always express the Floquet effective Hamiltonian $H(k)$
as 
\begin{equation}
	H(k)=h_{0}(k)\sigma_{0}+h_{x}(k)\sigma_{x}+h_{y}(k)\sigma_{y}+h_{z}(k)\sigma_{z}.\label{eq:Hk}
\end{equation}
Here $\sigma_{0}$ is the two by two identity matrix. $\sigma_{x}$,
$\sigma_{y}$ and $\sigma_{z}$ are Pauli matrices. $h_{0}(k)$, $h_{x}(k)$,
$h_{y}(k)$ and $h_{z}(k)$ can be real functions of $k$. 
By definition, the $H(k)$ here incorporates the all-round information of the system's stroboscopic dynamics over each complete driving period. Its quasienergy spectrum could thus be significantly different from and more complicated than the non-driven counterpart of the system.
By diagonalizing
$H(k)$, the dispersion relations of its two Floquet bands (defined
modulus $2\pi$ and indexed by $s$) are found to be
\begin{equation}
	E_{s}(k)=h_{0}(k)+sE(k),\qquad s=\pm,\label{eq:Esk}
\end{equation}
where
\begin{equation}
	E(k)=\sqrt{h_{x}^{2}(k)+h_{y}^{2}(k)+h_{z}^{2}(k)}.\label{eq:Ek}
\end{equation}
The associated Floquet eigenstates of $H(k)$ are further given by
\begin{equation}
	|\psi_{s}(k)\rangle=\frac{1}{\sqrt{2E(k)[E(k)+sh_{z}(k)]}}\begin{pmatrix}h_{z}(k)+sE(k)\\
		h_{x}(k)+ih_{y}(k)
	\end{pmatrix},\label{eq:Psisk}
\end{equation}
where $s=\pm$.
For any two eigenstates $|\psi_{s}(k)\rangle$ and $|\psi_{s}(k')\rangle$
of $H(k)$ in the same Floquet band, their overlap reads
\begin{widetext}
\begin{equation}
{\cal O}_{s}(k,k')\equiv\langle\psi_{s}(k)|\psi_{s}(k')\rangle=\frac{[h_{x}(k)-ih_{y}(k)][h_{x}(k')+ih_{y}(k')]+[E(k)+sh_{z}(k)][E(k')+sh_{z}(k')]}{2\sqrt{E(k)E(k')[E(k)+sh_{z}(k)][E(k')+sh_{z}(k')]}}.\label{eq:Ovlp}
\end{equation}
\end{widetext}
If $H(k)$ has the chiral symmetry ${\cal S}=\sigma_{z}$, such that
$\sigma_{z}H(k)\sigma_{z}=-H(k)$, we would have $h_{0}(k)=h_{z}(k)=0$
in Eq.~(\ref{eq:Hk}), and the overlap ${\cal O}_{s}(k,k')$ in 
Eq.~(\ref{eq:Ovlp}) reduces to
\begin{widetext}
\begin{equation}
	{\cal O}_{s}^{xy}(k,k')=\frac{1}{2}+\frac{[h_{x}(k)-ih_{y}(k)][h_{x}(k')+ih_{y}(k')]}{2E(k)E(k')}.\label{eq:OvlpXY}
\end{equation}
\end{widetext}
If $H(k)$ has the chiral symmetry ${\cal S}=\sigma_{y}$, such that
$\sigma_{y}H(k)\sigma_{y}=-H(k)$, we would have $h_{0}(k)=h_{y}(k)=0$
in Eq.~(\ref{eq:Hk}), and the overlap ${\cal O}_{s}(k,k')$ in 
Eq.~(\ref{eq:Ovlp}) reduces to
\begin{widetext}
\begin{equation}
	{\cal O}_{s}^{zx}(k,k')=\frac{h_{x}(k)h_{x}(k')+[E(k)+sh_{z}(k)][E(k')+sh_{z}(k')]}{2\sqrt{E(k)E(k')[E(k)+sh_{z}(k)][E(k')+sh_{z}(k')]}}.\label{eq:OvlpZX}
\end{equation}
\end{widetext}
Finally, when $H(k)$ has the chiral symmetry ${\cal S}=\sigma_{x}$
so that $\sigma_{x}H(k)\sigma_{x}=-H(k)$, we will have $h_{0}(k)=h_{x}(k)=0$
in Eq.~(\ref{eq:Hk}), and the overlap ${\cal O}_{s}(k,k')$ in 
Eq.~(\ref{eq:Ovlp}) becomes
\begin{widetext}
\begin{equation}
	{\cal O}_{s}^{yz}(k,k')=\frac{h_{y}(k)h_{y}(k')+[E(k)+sh_{z}(k)][E(k')+sh_{z}(k')]}{2\sqrt{E(k)E(k')[E(k)+sh_{z}(k)][E(k')+sh_{z}(k')]}}.\label{eq:OvlpYZ}
\end{equation}
\end{widetext}
The Eqs.~(\ref{eq:OvlpXY})--(\ref{eq:OvlpYZ}) will be used in the
calculations of quantum metric tensor and geometric entanglement entropy
for our 1D Floquet systems in the main text.

\section{QMT of 1D, two-band Hamiltonians\label{sec:QMT}}

In this Appendix, we deduce the quantum metric tensor of Floquet-Bloch
bands for 1D driven lattice models, with a focus on two-band settings.
For a 1D system described by the Floquet-Bloch effective Hamiltonian
in Eq.~(\ref{eq:Hk}), the quantum metric tensor \cite{QGT1980} in
$k$-space has a single component, i.e.,
\begin{alignat}{1}
	g_{kk}^s= & \langle\partial_{k}\psi_{s}|\partial_{k}\psi_{s}\rangle-\langle\partial_{k}\psi_{s}|\psi_{s}\rangle\langle\psi_{s}|\partial_{k}\psi_{s}\rangle\nonumber \\
	= & \langle\partial_{k}\psi_{s}|\psi_{-s}\rangle\langle\psi_{-s}|\partial_{k}\psi_{s}\rangle.\label{eq:gkk}
\end{alignat}
Here, $|\psi_{s}\rangle=|\psi_{s}(k)\rangle$ is given by Eq.~(\ref{eq:Psisk})
and $s=\pm$ label the two Floquet bands. According to Eq.~(\ref{eq:Ek}),
we could obtain
\begin{equation}
	\partial_{k}E(k)=\sum_{w=x,y,z}\frac{h_{w}(k)\partial_{k}h_{w}(k)}{E(k)},\label{eq:dE}
\end{equation}
and\textbf{
	\begin{equation}
		\partial_{k}\frac{1}{E(k)}=-\sum_{w=x,y,z}\frac{h_{w}(k)\partial_{k}h_{w}(k)}{E^{3}(k)}.\label{eq:dEinv}
	\end{equation}
}Using these relations together with the Eq.~(\ref{eq:Psisk}), we find
after straightforward calculations that
\begin{widetext}
\begin{equation}
	\langle\psi_{-s}|\partial_{k}\psi_{s}\rangle= \frac{s[h_{x}(h_{z}\partial_{k}h_{x}-h_{x}\partial_{k}h_{z})-h_{y}(h_{y}\partial_{k}h_{z}-h_{z}\partial_{k}h_{y})]+iE(h_{x}\partial_{k}h_{y}-h_{y}\partial_{k}h_{x})}{2E^{2}\sqrt{(E+h_{z})(E-h_{z})}}.
\end{equation}
\end{widetext}
Therefore, according to Eq.~(\ref{eq:gkk}), the QMT reads
\begin{widetext}
\begin{equation}
	g_{kk}=g_{kk}^s=\frac{[h_{x}(h_{z}\partial_{k}h_{x}-h_{x}\partial_{k}h_{z})-h_{y}(h_{y}\partial_{k}h_{z}-h_{z}\partial_{k}h_{y})]^{2}+E^{2}(h_{x}\partial_{k}h_{y}-h_{y}\partial_{k}h_{x})^{2}}{4E^{4}(E+h_{z})(E-h_{z})}.\label{eq:gkkH}
\end{equation}
\end{widetext}
It is clear that the $g_{kk}$ as obtained in Eq.~(\ref{eq:gkkH})
is independent of the Floquet band index $s$. 

Referring to the Appendix \ref{sec:Ovlp}, we have $h_{z}(k)=0$ for a
chiral symmetric $H(k)$ with ${\cal S}=\sigma_{z}$. The related
$g_{kk}$ then reduces to
\begin{equation}
	g_{kk}^{xy}=\frac{[h_{x}(k)\partial_{k}h_{y}(k)-h_{y}(k)\partial_{k}h_{x}(k)]^{2}}{4E^{4}(k)}.\label{eq:gkkXY}
\end{equation}
Similarly, for a chiral symmetric $H(k)$ with ${\cal S}=\sigma_{y}$,
we have $h_{y}(k)=0$ and the $g_{kk}$ reduces 
\begin{equation}
	g_{kk}^{zx}=\frac{[h_{z}(k)\partial_{k}h_{x}(k)-h_{x}(k)\partial_{k}h_{z}(k)]^{2}}{4E^{4}(k)}.\label{eq:gkkZX}
\end{equation}
Meanwhile, for an $H(k)$ with the chiral symmetry ${\cal S}=\sigma_{x}$,
we have $h_{x}(k)=0$ and the resulting QMT reads
\begin{equation}
	g_{kk}^{yz}=\frac{[h_{y}(k)\partial_{k}h_{z}(k)-h_{z}(k)\partial_{k}h_{y}(k)]^{2}}{4E^{4}(k)}.\label{eq:gkkYZ}
\end{equation}
Note in passing that in all the cases we have $g_{kk}\geq0$, as expected.
With the help of Eqs.~(\ref{eq:gkkXY})--(\ref{eq:gkkYZ}), we could further
work out the integration of QMT for the different Floquet models considered
in the main text.

For 1D chiral symmetric models, a connection between the QMT and the
topological winding number could be identified. Let us denote the effective
Floquet-Bloch Hamiltonian of such a system as $H_{ab}(k)=h_{a}(k)\sigma_{a}+h_{b}(k)\sigma_{b}$,
where $a,b=x,y,z$ and $a\neq b$. The chiral symmetry operator of
$H_{ab}(k)$ is thus the Pauli matrix $\sigma_{c}$ with $c=x,y,z$
and $c\neq a,b$. The topological phases of the system described by
$H_{ab}(k)$ can be characterized by the integer winding number
\begin{equation}
	w=\int_{-\pi}^{\pi}\frac{dk}{2\pi}\partial_{k}\phi^{ab}(k),\label{eq:w}
\end{equation}
where the winding angle $\phi^{ab}(k)\equiv\arctan[h_{b}(k)/h_{a}(k)]$,
and thus
\begin{equation}
	\partial_{k}\phi^{ab}(k)=\frac{h_{a}(k)\partial_{k}h_{b}(k)-h_{b}(k)\partial_{k}h_{a}(k)}{E^{2}(k)},\label{eq:dphi}
\end{equation}
with $E^{2}(k)=h_{a}^{2}(k)+h_{b}^{2}(k)$. According to Eqs.~(\ref{eq:gkkXY})--(\ref{eq:gkkYZ}),
the QMT of such a system reads
\begin{equation}
	g_{kk}^{ab}=\frac{[h_{a}(k)\partial_{k}h_{b}(k)-h_{b}(k)\partial_{k}h_{a}(k)]^{2}}{4E^{4}(k)}.\label{eq:gkkAB}
\end{equation}
We then arrive at the relation
\begin{equation}
	g_{kk}^{ab}=\frac{1}{4}[\partial_{k}\phi^{ab}(k)]^{2}.\label{eq:gkkphik}
\end{equation}
This equation allows us to obtain the QMT from the topological winding
angle for a two-band Floquet-Bloch Hamiltonian with chiral symmetry
in one dimension. It unveils an interesting connection between the
quantum geometry and topology in 1D systems, which is different from
the case reflected in the Zak phase.

\section{EE of Floquet states\label{sec:EE}}

In this Appendix, we describe an approach to obtain the bipartite
von Neumann and R\'enyi EE for a many-particle state of noninteracting
fermions with an arbitrary filling fraction over a Floquet band in 1D systems.

We start with the general relationship between the single-particle
correlation matrix and the reduced density matrix of a bipartite system,
which was well-established for static free lattice models \cite{ESEERev04}
and generalized also to Floquet models recently \cite{LWZPRR2022}.
Let us consider a noninteracting many-particle system $S$ prepared
in the pure state $|\Psi\rangle$ and a subsystem $A$ belonging to
$S$. We can obtain the reduced density matrix of $A$ as $\hat{\rho}_{A}={\rm Tr}_{\overline{A}}(\hat{\rho})$.
Here $\hat{\rho}=|\Psi\rangle\langle\Psi|$ is the density matrix
of whole system $S=A\cup\overline{A}$. The trace ${\rm Tr}_{\overline{A}}$
is taken over the degrees of freedom belonging to the subsystem $\overline{A}$
complementing to $A$. If $|\Psi\rangle$ represents a Gaussian state,
we could always write \cite{ESEERev04}
\begin{equation}
	\hat{\rho}_{A}=\frac{1}{Z}e^{-\hat{H}_{A}},\qquad Z\equiv{\rm Tr}(e^{-\hat{H}_{A}}),\label{eq:RhoA}
\end{equation}
where $Z$ is a normalization factor and $\hat{H}_{A}$ is usually
called the entanglement Hamiltonian \cite{ESEERev04}. Any single-particle
correlation function restricted to the subsystem $A$ can now be evaluated
as
\begin{equation}
	C_{m,n}^{A}={\rm Tr}(\hat{c}_{m}^{\dagger}\hat{c}_{n}\hat{\rho}_{A})=\frac{{\rm Tr}(\hat{c}_{m}^{\dagger}\hat{c}_{n}e^{-\hat{H}_{A}})}{{\rm Tr}(e^{-\hat{H}_{A}})},\label{eq:CmnA}
\end{equation}
where $\{m,n\}\in A$ and $\hat{c}_{m}^{\dagger}$ ($\hat{c}_{n}$)
creates (annihilates) a fermion into (from) the single-particle state
$|m\rangle$ ($|n\rangle$) inside the subsystem $A$. 

Let $\{|\phi_{j}\rangle|j\in A\}$ be the complete and orthonormal
eigenbasis of $\hat{H}_{A}$ with the eigenvalues $\{\xi_{j}\}$,
such that the entanglement Hamiltonian admits the spectral decomposition
\begin{equation}
	\hat{H}_{A}=\sum_{j\in A}\xi_{j}\hat{\phi}_{j}^{\dagger}\hat{\phi}_{j}.\label{eq:HA}
\end{equation}
Here the set $\{\xi_{j}\}$ is usually referred to as the entanglement
spectrum (ES) \cite{ESPRL2008} of subsystem $A$. Since both $\{|n\rangle|n\in A\}$
and $\{|\phi_{j}\rangle|j\in A\}$ form normalized bases of the subsystem
$A$, their corresponding creation and annihilation operators can
be related to each other by unitary transformations, i.e.,
\begin{equation}
	\hat{c}_{n}^{\dagger}=\sum_{j\in A}\phi_{nj}^{*}\hat{\phi}_{j}^{\dagger},\qquad\hat{c}_{n}=\sum_{j\in A}\phi_{nj}\hat{\phi}_{j},\label{eq:cphi}
\end{equation}
where $\phi_{nj}=\langle n|\phi_{j}\rangle$. Plugging Eqs.~(\ref{eq:HA})
and (\ref{eq:cphi}) into Eq.~(\ref{eq:CmnA}) and carrying out straightforward
calculations, we could re-express the $C_{m,n}^{A}$ as \cite{ESEERev04}
\begin{equation}
	C_{m,n}^{A}=\sum_{j\in A}\frac{\langle n|\phi_{j}\rangle\langle\phi_{j}|m\rangle}{e^{\xi_{j}}+1}.\label{eq:CmnA2}
\end{equation}
Therefore, the single-particle correlation matrix $C^{A}$ admits
the spectral decomposition
\begin{equation}
	(C^{A})^{\top}=\sum_{j\in A}\zeta_{j}|\phi_{j}\rangle\langle\phi_{j}|,\qquad\zeta_{j}=\frac{1}{e^{\xi_{j}}+1}.\label{eq:CAT}
\end{equation}
It is now clear that there is a one-to-one correspondence between
the spectrum $\{\zeta_{j}\}$ of $C^{A}$ and the ES $\{\xi_{j}\}$
of $\hat{H}_{A}$, i.e.,
\begin{equation}
	\xi_{j}=\ln(\zeta_{j}^{-1}-1).\label{eq:ES}
\end{equation}
This relation allows us to deduce the ES, EE and the related quantities
such as mutual information of a bipartite system from the spectrum
of its single-particle correlation matrix \cite{ESEERev04}.

Next, we try to rewrite the reduced density matrix $\hat{\rho}_{A}$
in terms of $C^{A}$, which allows us to obtain the EE directly from
the spectrum of correlation matrix. From Eqs.~(\ref{eq:RhoA}), (\ref{eq:HA}),
(\ref{eq:CAT}) and (\ref{eq:ES}), we find
\begin{equation}
	\frac{1}{Z}=\prod_{j\in A}(1-\zeta_{j})=\det(\mathbb{I}^{A}-C^{A}),\label{eq:1ovZ}
\end{equation}
where $\mathbb{I}^{A}$ is the identity matrix of subsystem $A$.
Using the inverse of the transformations between different bases in
Eq.~(\ref{eq:cphi}), i.e.,
\begin{equation}
	\hat{\phi}_{j}^{\dagger}=\sum_{n\in A}\phi_{nj}\hat{c}_{n}^{\dagger},\qquad\hat{\phi}_{j}=\sum_{n\in A}\phi_{nj}^{*}\hat{c}_{n},\label{eq:phic}
\end{equation}
we could further obtain from Eqs. (\ref{eq:RhoA}), (\ref{eq:HA}) and (\ref{eq:CAT})
that
\begin{equation}
	e^{-\hat{H}_{A}}=e^{-\sum_{m,n}\ln[(C^{A})^{-1}-\mathbb{I}^{A}]_{m,n}^{\top}\hat{c}_{m}^{\dagger}\hat{c}_{n}}.\label{eq:e-HA}
\end{equation}
Putting together, we find the expression of $\hat{\rho}_{A}$ in terms
of $C^{A}$ as
\begin{equation}
	\hat{\rho}_{A}=\det(\mathbb{I}^{A}-C^{A})e^{-\sum_{m,n\in A}\ln[(C^{A})^{-1}-\mathbb{I}^{A}]_{m,n}^{\top}\hat{c}_{m}^{\dagger}\hat{c}_{n}}.\label{eq:rhoACA}
\end{equation}
Equivalently, in terms of the eigenvalues of $C^{A}$, we would have

\begin{equation}
	\hat{\rho}_{A}=\left[\prod_{j\in A}(1-\zeta_{j})\right]e^{-\sum_{j\in A}\ln(\zeta_{j}^{-1}-1)\hat{\phi}_{j}^{\dagger}\hat{\phi}_{j}}.\label{eq:rhoAzetaj}
\end{equation}

For a bipartite system $S=A\cup\overline{A}$, the $\lambda$th R\'enyi
EE and von Neumann EE between $A$ and $\overline{A}$ are defined
as
\begin{equation}
	S_{A}^{(\lambda)}\equiv\frac{1}{1-\lambda}\ln{\rm Tr}\hat{\rho}_{A}^{\lambda},\label{eq:REE}
\end{equation}
\begin{equation}
	S_{A}\equiv-{\rm Tr}(\hat{\rho}_{A}\ln\hat{\rho}_{A})=\lim_{\lambda\rightarrow1}S_{A}^{(\lambda)}.\label{eq:vNEE}
\end{equation}
Plugging Eq.~(\ref{eq:rhoAzetaj}) into Eqs.~(\ref{eq:REE}) and (\ref{eq:vNEE}),
we could directly find
\begin{equation}
	S_{A}^{(\lambda)}=\frac{1}{1-\lambda}\sum_{j\in A}\ln[\zeta_{j}^{\lambda}+(1-\zeta_{j})^{\lambda}],\label{eq:REE2}
\end{equation}
\begin{equation}
	S_{A}=-\sum_{j\in A}[\zeta_{j}\ln\zeta_{j}+(1-\zeta_{j})\ln(1-\zeta_{j})].\label{eq:vNEE2}
\end{equation}

Therefore, the spectrum $\{\zeta_{j}\}$ of single-particle correlation
matrix $C^{A}$ could provide us with complete information about the
bipartite EE between the subsystem $A$ and its complement $\overline{A}$
for a given multi-particle Gaussian state $|\Psi\rangle$ of the whole
system $S$. Note in passing that the relations in Eqs.~(\ref{eq:REE2})
and (\ref{eq:vNEE2}) are applicable to both static and Floquet systems
made up of noninteracting fermions in Gaussian states \cite{LWZPRR2022}.
For a Floquet system with a one-period evolution (Floquet) operator
$\hat{U}$, one could start with the multi-particle state in the form
of $|\Psi\rangle=\prod_{\ell\in{\rm occ.}}\hat{\psi}_{\ell}^{\dagger}|\emptyset\rangle$
and the resulting density operator $\hat{\rho}=|\Psi\rangle\langle\Psi|$,
where $|\emptyset\rangle$ is the vacuum state and $\hat{\psi}_{\ell}^{\dagger}$
creates a fermion in the single-particle Floquet eigenbasis $|\psi_{\ell}\rangle$
of $\hat{U}$ \cite{LWZPRR2022}. 

\section{EE and the overlap matrix\label{sec:EEvsOA}}

In this Appendix, we discuss an approach to obtain the EE from the
overlap matrix restricted to a given subsystem \cite{Fredholm01,Fredholm02,Fredholm03},
which encodes the quantum geometry of the latter \cite{PaulPRB2024}.
We start with the Fredholm determinant
\begin{equation}
	D^{A}(\zeta)\equiv\det(\zeta\mathbb{I}^{A}-C^{A})=\prod_{j\in A}(\zeta-\zeta_{j}).\label{eq:DA}
\end{equation}
Here the meanings of $\mathbb{I}^{A}$, $C^{A}$ and $\{\zeta_{j}\}$
for subsystem $A$ are the same as those introduced in Appendix \ref{sec:EE}.
Note in passing that the correlation-matrix eigenvalue $\zeta_{j}$
has the range $[0,1]$ for any $j$. In terms of the Fredholm determinant,
we could express the R\'enyi EE $S_{A}^{(\lambda)}$ in terms of a contour
integration encircling the segment $[0,1]$ of the real-axis, i.e.,
\begin{equation}
	S_{A}^{(\lambda)}=\oint\frac{d\zeta}{2\pi i}\frac{1}{1-\lambda}\ln[\zeta^{\lambda}+(1-\zeta)^{\lambda}]\frac{d\ln D^{A}(\zeta)}{d\zeta}.\label{eq:REE3}
\end{equation}
The related von Neumann EE can further be obtained by taking the limit
$\lambda\rightarrow1$. 

For any two occupied single-particle states $|\psi_{\ell}\rangle$
and $|\psi_{\ell'}\rangle$ in a composite system $S=A\cup\overline{A}$,
their overlap within the subsystem $A$ can be defined as
\begin{equation}
	O_{\ell,\ell'}^{A}=\sum_{n\in A}\langle\psi_{\ell}|n\rangle\langle n|\psi_{\ell'}\rangle=\sum_{n\in A}\psi_{n\ell}^{*}\psi_{n\ell'}.\label{eq:OllA}
\end{equation}
If we have in total $N$ such occupied states $\{|\psi_{\ell}\rangle|\ell=1,...,N\}$,
all the quantum geometry of this state manifold that are associated
to the subsystem $A$ should be captured by the $N\times N$ overlap
matrix $O^{A}$ with elements $\{O_{\ell,\ell'}^{A}|\ell,\ell'=1,...,N\}$,
which are given by Eq.~(\ref{eq:OllA}).

We could now establish a connection between the spectra of the overlap
matrix $O^{A}$ and the single-particle correlation matrix $C^{A}$
{[}Eq.~(\ref{eq:CmnA}){]} \cite{Fredholm01,Fredholm02,Fredholm03},
which further allows us to figure out the quantum-geometric component
of EE. Let us consider the $q$th power of $O^{A}$, whose trace is
given by
\begin{widetext}
\begin{equation}
	{\rm Tr}[(O^{A})^{q}]= \sum_{n_{1},...,n_{q}}\sum_{\ell_{1},...,\ell_{q}}\langle\psi_{\ell_{1}}|n_{1}\rangle\langle n_{1}|\psi_{\ell_{2}}\rangle\langle\psi_{\ell_{2}}|n_{2}\rangle\langle n_{2}|\psi_{\ell_{3}}\rangle
	\cdots\langle\psi_{\ell_{q-1}}|n_{q-1}\rangle\langle n_{q-1}|\psi_{\ell_{q}}\rangle\langle\psi_{\ell_{q}}|n_{q}\rangle\langle n_{q}|\psi_{\ell_{1}}\rangle.\label{eq:TrOAq}
\end{equation}
\end{widetext}
For the $N$-particle state $|\Psi\rangle=\prod_{\ell=1}^{N}\hat{\psi}_{\ell}^{\dagger}|\emptyset\rangle$
and for any $\ell\in\ell_{1},...,\ell_{q}$, we have 
\begin{equation}
	\sum_{\ell}\langle\psi_{\ell}|m\rangle\langle n|\psi_{\ell}\rangle=\langle\Psi|\hat{c}_{m}^{\dagger}\hat{c}_{n}|\Psi\rangle=C_{m,n}^{A},\label{eq:CmnA3}
\end{equation}
where $m,n\in A$ and $C_{m,n}^{A}$ is the correlation-matrix element
of subsystem $A$ {[}Eq.~(\ref{eq:CmnA}){]}. Inserting Eq.~(\ref{eq:CmnA3})
into Eq.~(\ref{eq:TrOAq}), we find [after reorganizing the terms in
Eq.~(\ref{eq:TrOAq})] that
\begin{equation}
	{\rm Tr}[(O^{A})^{q}]={\rm Tr}[(C^{A})^{q}].\label{eq:TrOACA}
\end{equation}
Therefore, taking any power $q\in\mathbb{N}$, the trace of the overlap
matrix $O^{A}$ and the correlation matrix $C^{A}$ restricted to the
subsystem $A$ are identical. Such a connection would allow us to
express EE in terms of the eigenvalues of $O^{A}$, within which the
quantum geometric properties of the occupied states $\{|\psi_{\ell}\rangle|\ell=1,...,N\}$
are encoded.

To proceed, we take the logarithm of the Fredholm determinant in 
Eq.~(\ref{eq:DA}), yielding \cite{Fredholm01,Fredholm02,Fredholm03}
\begin{equation}
	\ln[D^{A}(\zeta)]=\sum_{j}\ln(\zeta-\zeta_{j})=\sum_{j}\left(\ln\zeta-\sum_{q=1}^{\infty}\frac{\zeta_{j}^{q}}{q\zeta^{q}}\right).\label{eq:LnDA}
\end{equation}
Let $\{\eta_{\ell}|\ell=1,...,N\}$ be the eigenvalues of the overlap
matrix $O^{A}$, we obtain from Eq.~(\ref{eq:TrOACA}) that
\begin{equation}
	\sum_{\ell}\eta_{\ell}^{q}={\rm Tr}[(O^{A})^{q}]={\rm Tr}[(C^{A})^{q}]=\sum_{j}\zeta_{j}^{q}.\label{eq:OACA}
\end{equation}
Combining Eqs.~(\ref{eq:LnDA}) and (\ref{eq:OACA}) into Eq.~(\ref{eq:REE3})
finally leads us to another explicit expression for the R\'enyi bipartite
EE $S_{A}^{(\lambda)}$, i.e.,
\begin{equation}
	S_{A}^{(\lambda)}=\frac{1}{1-\lambda}\sum_{\ell=1}^{N}\ln\left[\eta_{\ell}^{\lambda}+(1-\eta_{\ell})^{\lambda}\right],\label{eq:REE4}
\end{equation}
and also for the von Neumann EE
\begin{equation}
	S_{A}=-\sum_{\ell=1}^{N}\left[\eta_{\ell}\ln\eta_{\ell}+(1-\eta_{\ell})\ln(1-\eta_{\ell})\right].\label{eq:vNEE4}
\end{equation}
Note in passing that $N$ here counts the total number of occupied single-particle
states, which is fixed regardless of the size of subsystem $A$. In
summary, for a given group of occupied single-particle states $\{|\psi_{\ell}\rangle|\ell=1,...,N\}$
within a noninteracting fermionic system, we can obtain the bipartite
EE through Eqs.~(\ref{eq:REE4}) and (\ref{eq:vNEE4}) after getting
the eigenspectrum $\{\eta_{\ell}|\ell=1,...,N\}$ of the overlap matrix
$O^{A}$ {[}Eq.~(\ref{eq:OllA}){]}. Since $O^{A}$ encodes the quantum
geometry of many-particle state $|\Psi\rangle=\prod_{\ell}|\psi_{\ell}\rangle$,
we expect to identify geometric contributions to EE from Eqs.~(\ref{eq:REE4})
and (\ref{eq:vNEE4}) after removing possible non-geometric components
\cite{PaulPRB2024}. It deserves to mention that the results developed
here could be equally applicable to both static and Floquet systems.
For the latter case, we simply regard $\{|\psi_{\ell}\rangle|\ell=1,...,N\}$
as a set of occupied single-particle Floquet eigenstates of a periodically
driven quantum system.

\section{GEE of Floquet states\label{sec:GEE}}

In this Appendix, we discuss a scheme of decomposing EE into a geometric
part (GEE) and a non-geometric contribution following Ref.~\cite{PaulPRB2024}.
We restrict our attention to the quantum geometry of 1D systems in
wave-vector space. The formalism discussed here are not hard to be
generalized to higher spatial dimensions and to other kinds of parameter
spaces.

Let $\{|\varphi_{k}\rangle\}$ be a set of eigenstates populating
a single Floquet-Bloch band (under PBC), the overlap matrix element
in Eq.~(\ref{eq:OllA}) can be expressed in this case as
\begin{equation}
	O_{k,k'}^{A}=\sum_{n\in A}\langle\varphi_{k}|n\rangle\langle n|\varphi_{k'}\rangle.\label{eq:OAkkp}
\end{equation}
Assuming that there are $L$ unit cells in the 1D lattice and each
unit cell has $p$ internal degrees of freedom (spins, sublattices,
etc.), the wave function overlap takes the form
\begin{equation}
	\langle n|\varphi_{k}\rangle=\frac{1}{\sqrt{L}}e^{ikn}\begin{pmatrix}b_{1}(k)\\
		\vdots\\
		b_{p}(k)
	\end{pmatrix}\equiv\frac{1}{\sqrt{L}}e^{ikn}|\psi_{k}\rangle.\label{eq:npsik}
\end{equation}
It allows us to re-express the $O_{k,k'}^{A}$ in Eq.~(\ref{eq:OAkkp})
as
\begin{equation}
	O_{k,k'}^{A}=\frac{1}{L}\sum_{n\in A}e^{-i(k-k')n}\langle\psi_{k}|\psi_{k'}\rangle.\label{eq:OAkkp2}
\end{equation}
For a subsystem $A$ with $L_{A}$ unit cells, the summation in 
Eq.~(\ref{eq:OAkkp2}) can be worked out, yielding
\begin{equation}
	O_{k,k'}^{A}=O_{k,k'}^{A_{0}}\langle\psi_{k}|\psi_{k'}\rangle,\label{eq:OAkkp3}
\end{equation}
where
\begin{equation}
	O_{k,k'}^{A_{0}}=\begin{cases}
		L_{A}/L & k=k',\\
		\frac{\sin[(k-k')L_{A}/2]}{L\sin[(k-k')/2]}e^{\frac{i}{2}(k-k')} & k\neq k'.
	\end{cases}\label{eq:OAkkp0}
\end{equation}
The coefficient $O_{k,k'}^{A_{0}}$ is generic and it describes the
overlap matrix element of a single-band lattice model in one-dimension, whose
related quantum geometry is trivial in $k$-space. Therefore, if we
remove the contributions from the spectrum of $O^{A_{0}}$ to EE, we
will be left with the part of EE that is originated from the quantum
geometry of a multi-band system. Based on this understanding, we may define
the R\'enyi GEE between two subsystems $A$ and $\overline{A}$ as
\begin{equation}
	S_{{\rm QG}}^{(\lambda)}=S_{A}^{(\lambda)}-S_{A_{0}}^{(\lambda)},\label{eq:RGEE}
\end{equation}
where $S_{A}^{(\lambda)}$ and $S_{A_{0}}^{(\lambda)}$ are obtained
by inserting the spectrum of $O^{A}$ {[}Eq.~(\ref{eq:OAkkp3}){]}
and $O^{A_{0}}$ {[}Eq.~(\ref{eq:OAkkp0}){]} into Eq.~(\ref{eq:REE4}),
respectively. The von Neumann EE then reads
\begin{equation}
	S_{{\rm QG}}=S_{A}-S_{A_{0}}.\label{eq:vNGEE}
\end{equation}
Physically, the GEE defined here characterizes the EE due to multi-band
quantum geometric effects \cite{PaulPRB2024}. It omits the contribution
from a set of fermions with the same population as the multi-band
system but with trivial quantum geometries. We will use the Eq.~(\ref{eq:vNGEE})
to describe the GEE of different Floquet models considered in the
main text.

\end{document}